\documentclass[twocolumn,aps,pra,amssymb,showpacs,longbibliography,10pt]{revtex4-1}

\usepackage{amsfonts,amssymb,amsmath,graphicx,color,hyperref,extarrows}
\usepackage{ulem}

\usepackage{multirow}
\newcommand{\br}{{\bf r}}
\newcommand{\vq}{{\bf r}}

\newcommand{\vk}{{\bf k}}

\newcommand{\echo}{\text{echo}}

\newcommand{\peak}{\text{peak}}
\newcommand{\vsigma}{\boldsymbol{\sigma}}
\newcommand{\reqA}{a)}
\newcommand{\reqB}{b)}
\newcommand{\reqC}{c)}
\newcommand{\reqD}{d)}
\newcommand{\casepertA}{A)}
\newcommand{\casepertB}{B)}
\newcommand{\casepertC}{C)}
\usepackage{bbm} %1-Matrix

\newcommand{\pulse}{\text{pulse}}
\newcommand{\imp}{\text{imp}}

\newcommand{\tref}{\text{ref}}

\def\be{\begin{equation}}
\def\ee{\end{equation}}
\def\ber{\begin{eqnarray}}
\def\eer{\end{eqnarray}}
\def\nn{\nonumber}
\newcommand{\bh}{{\bf h}}
\newcommand{\Mcal}{{\mathcal M}}
\newcommand{\hbm}{\hat{\bf m}}
\newcommand{\hbn}{\hat{\bf n}}
\newcommand{\hbl}{\hat{\bf l}}
\newcommand{\Ecal}{{\mathcal E}}
\newcommand{\Ecalb}{{\boldsymbol{\mathcal E}}}

\newcommand{\iu}{{i\mkern1mu}} 	%imaginary unit
\newcommand{\me}{\mathrm{e}} 	%euler
\newcommand{\ie}{{\it i.e.~}} 	%i.e.
\newcommand{\eg}{{\it e.g.~}} 	%e.g.
 
\definecolor{greenPR}{rgb}{0.00, 0.6, 0.00}
\newcommand{\commentout}[1]{}

\graphicspath{{./pictures/}}

\begin{document}

\title{Quantum time mirrors for general two-band systems}
\author{Phillipp Reck$^{1,2}$, Cosimo Gorini$^1$, Klaus Richter$^{1}$}
\email[]{klaus.richter@ur.de}
\affiliation{$^1$Institut f\"ur Theoretische Physik, Universit\"at Regensburg, 93040 Regensburg, Germany}
\affiliation{$^2$SPEC, CEA, CNRS, Université Paris-Saclay, CEA Saclay, 91191 Gif-sur-Yvette Cedex, France}
\date{\today}

\begin{abstract}
Methods that are devised to achieve reversal of quantum dynamics in time have been named ``quatum time mirrors''.
Such a time mirror can be considered as a generalization of Hahn's spin echo to systems with continuous degrees of freedom.
We extend the quantum time mirror protocol originally proposed for Dirac dispersions to arbitrary two-band systems and establish the general requirements for its efficient implementation.
We further discuss its sensitivity to various non-homogeneous perturbations including disorder potentials and the effect of external static magnetic and electric fields.   
Our general statements are verified for a number of exemplary Hamiltonians, 
whose phase-coherent dynamics are studied both analytically and numerically.
The Hamiltonians considered can be used to describe the low-energy properties of systems as diverse as cold atom-optics setups,
direct band gap semiconductors or (mono- or bilayer) graphene.  We discuss the consequences of many-body effects at a qualitative level,
and consider the protocol feasibility in state-of-the-art experimental setups.
\end{abstract}

\maketitle

%%%%%%%%%%%%%%%%%%%%%%%%%%%%%%%% Intro %%%%%%%%%%%%%%%%%%%%%%%%%%%%%%%%%%%%%%%%%%%%%%%%%%%%

\section{Introduction}
\label{sec_intro}
After a slow start, the physics of time-reversal made rapid progress during the last 30 years. 
While physicists like Boltzmann \cite{boltzmann1877} and Loschmidt \cite{loschmidt1876} were arguing about time-reversal
and entropy increase already in the 19th century, and Eddington \cite{eddingtonbook} later coined the expression ``arrow of time'' in this context, 
the only time-reversal protocol realized -- and later technically exploited -- before the 1990s was Hahn's spin echo, dating back to 1950 \cite{hahn1950}.
Its technical developments and benefits, \eg in noninvasive imaging of biological tissues, are impressive
\cite{zhang2012,zoehrer2017}.
The pace increased in the 1990s with a series of novel time-reversal protocols, so-called ``time-reversal mirrors''.
Such time mirrors usually follow a record-and-play back scheme: the signal from an initial wave, scattered while traversing a random medium,
is recorded by an array of receiver-emitter antennas positioned around the scattering region.  At a later time
the signal is rebroadcast in a time-reversed fashion, \ie what came in last goes out first, and is refocused by the 
random medium approximately recreating the original input wave at the source.  
Protocols of this kind were developed for ultrasonic \cite{fink1992}, elastic \cite{fink1997}, 
acoustic \cite{draeger1997,fink1999}, electromagnetic \cite{lerosey2004} and, very recently, water waves \cite{przadka2012,
chabchoub2014}.  They are now employed in medicine, material analysis,
telecommunication and other fields were wave control is of critical importance \cite{fink1997,fink1999,lerosey2007,mosk2012}.

Their implementation for visible light was complicated by the lack of available antennas, 
and nonlinear effects like three- and four-wave mixing were used for time-reversal purposes \cite{yariv1978,miller1980}. 
In 2011, Sivan and Pendry proposed to circumvent this problem with the use of photonic crystals,
whose spatio-temporal modulation leads to an inversion of the light propagation direction \cite{sivan2011a,sivan2011b,sivan2011c}.

The concept of a record-and-play back mirror is however not directly transferable to quantum systems, 
due to the nature of the measurement process.  Even if theoretical scenarios for non-invasive quantum 
detectors and emitters were investigated \cite{pastawski2007,calvo2010}, various time-reversal protocols based on alternative
concepts were proposed for quantum systems over the past ten years, 
\eg for the kicked rotator simulated for a Bose-Einstein condensate \cite{martin2008,ullah2011} or 
for magnonic crystals \cite{chumak2010,karenowska2012}.
In the kicked rotator setup, the time-reversal is induced by a change of the kicking protocol and is feasible only in a small momentum range. 
Magnonic crystals rely instead on diabatically switching on and off a spatially periodic potential, which couples certain modes with different propagation directions.

Recently an ``instantaneous time mirror'', \ie a new form of a classical time mirror circumventing the measurement process, 
was implemented \cite{bacot2016}.  Here the propagation of gravity-capillary water waves is inverted by a short 
and homogeneous perturbation, specifically a vertical acceleration of the water bath.
The resulting generation of Cauchy sources cause parts of the wave to move back to their initial position, 
creating an echo of the original signal.  Besides not requiring any recording, such an homogeneous, instantaneous time mirror
has an additional advantage: it is effective nearly independent of the shape and the position of the initial wave.

%%%%%%%%%%%%%%%%%%%%%%%%%%%%% FIGURE %%%%%%%%%%%%%%%%%%%%%%%%%%%%%%%%%%%%%%%%%%%%%
\begin{figure}
 \centering 

     \includegraphics[width=\columnwidth]{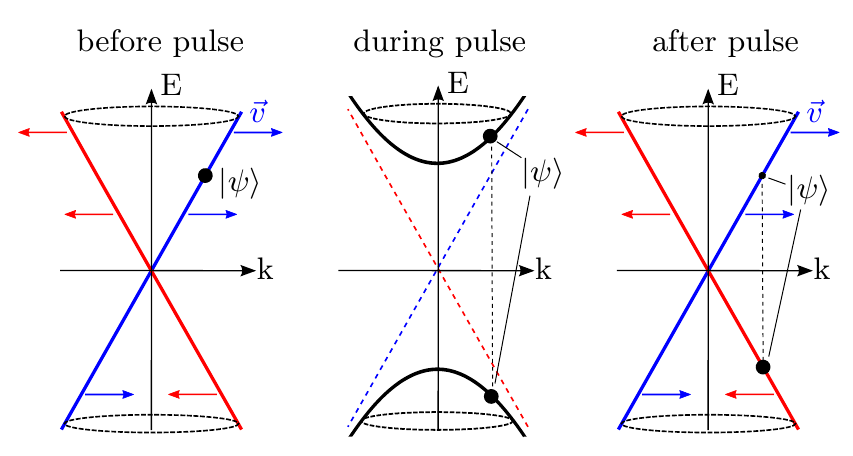}
\caption{
Mechanism of the Dirac Quantum Time Mirror (QTM) as used in Ref.~\cite{reck2017} and extended here to general two-band dispersions. 
By opening a gap in the beforehand linear band structure, an initial eigenstate (blue band) is split into a superposition of the upper and lower band and undergoes Rabi-like oscillations.
Depending on the band opening (duration and gap energy) it will end up mainly in the other (red) band after switching off the gap pulse.
Since the group velocity is exactly opposite in the upper and lower band as indicated by the colored arrows, the propagation of the state is inverted and thus effectively time-inverted.
}\label{fig:QTM-mechanism}
\end{figure}
%%%%%%%%%%%%%%%%%%%%%%%%%%%%% FIGURE %%%%%%%%%%%%%%%%%%%%%%%%%%%%%%%%%%%%%%%%%%%%%

Instantaneous time-reversal protocols for quantum systems based on fast homogeneous pulses 
were recently proposed \cite{reck2017,reck2018}.  The non-linear protocol of Ref.~\cite{reck2018} and generalizations \cite{goussev2018} are designed for
systems obeying a non-linear Schr\"odinger equation, such as atomic Bose-Einstein condensates in optical traps.
The Quantum Time Mirror (QTM) of Ref.~\cite{reck2017} was instead put forward for Dirac systems,
and relies on pulse-induced oscillations between upper and lower branches of the Dirac cone, see Fig.~\ref{fig:QTM-mechanism}. 
Since each branch is characterized by velocities of equal magnitude but opposite sign, a transition tuned 
such as to convert an initially upper (lower) branch state into the lower (upper) branch ideally achieves perfect velocity reversal,
leading to an effective propagation ``back in time''.
The principle is a generalization of Hahn's spin echo mechanism: The latter is based on coherent oscillations in a discrete two-level system,
while the QTM mechanism exploits analogous oscillations in continuous two-band states.
Though there are some advantages in linear dispersions as discussed below, it is not limited to them,
and the purpose of this paper is to extend the QTM to general two-band systems.
The goal of our in-depth study is to identify the general requirements for efficient time-reversal
and to characterize their relative importance and sensitivity to perturbations in non-ideal settings.
Within a single and coherent (quasi-)particle picture, the approach we follow is rather general, 
in particular not tied to certain materials or (quasi-)particle types, \ie fermionic or bosonic.
We refer to ``electron-'' and ``hole-like'' states to label the upper and lower branch
of an abstract two-band system of unspecified nature.
% Many-body effects due to interactions or \eg the presence of a Fermi sea will be discussed qualitatively in the closing.

The paper is organized as follows.
In Sec.~\ref{sec:genmod}, we discuss the general features that a system needs to fulfill in order to qualify as a QTM setup.
The requirements for a good echo are formulated from a general standpoint
and summarized in Table \ref{table_reqs}.
They are then examined in Sec.~\ref{sec:twoband-examples} for four classes of systems: 
(i) a perfect Dirac cone (ii) two linear but asymmetric bands (iii) hyperbolic and (iv) parabolic bands with different curvatures.
The effects of perturbations to the ideal QTM Hamiltonian, in the sense of additional terms \eg due to disorder or external
electromagnetic fields, are discussed in Sec.~\ref{sec:Perturbations}: first in general terms (Subsec.~\ref{subsec:pert:gen}), 
and later for static electric and magnetic fields (Subsecs.~\ref{subsec:electric} and \ref{subsec:magnetic}).
The role of disorder was extensively investigated in Ref.~\cite{reck2017}, and will only be briefly recalled here.
We qualitatively address many-body physics in Sec.~\ref{sec:manybody}, and sum up our results in Sec.~\ref{sec_conclusions}.
Details of certain calculations are found in the Appendices.

%%%%%%%%%%%%%%%%%%%%%%%%%%%%%%%% Graphene %%%%%%%%%%%%%%%%%%%%%%%%%%%%%%%%%%%%%%%%%%%%%%%%%%%%

\section{General model}
\label{sec:genmod}

\subsection{QTM velocity requirements}
\label{subsec:genvelreq}

Take a general two-band system subjected to an homogeneous pulse which diabatically couples the bands.
Such a pulse is wavevector conserving, so that
any chiral symmetric dispersion could qualify as an efficient QTM: Vertical transitions between
electron- and hole-like branches imply exact reversal of the velocity at that given wavevector value.
\footnote{A dispersion is chiral symmetric if its particle- ($E>0$) and hole-like ($E<0$) branches 
are the mirror image of each other with respect to the $E=0$ line.}
Actually, the requirement of chiral symmetry can be relaxed, as it is sufficient for the two branches
to be proportional to each other.  We quantify this for two bands $E^\pm_\vk$ and the corresponding group velocities 
\begin{equation}
{\bf v}^\pm_\vk = \nabla_\vk E^\pm_\vk.
\end{equation}
For chiral symmetric linear bands, as considered in Ref.~\cite{reck2017}, the magnitude of the velocity is $\vk$-independent 
and its direction is (anti)parallel to $\vk$ for positive (negative) energies. 
Therefore a pulse-induced transition $E_\vk^\pm \rightarrow E_\vk^\mp$ ensures that 
all $\vk$-modes propagate back and reach the origin at the same time.
This yields a high fidelity echo at $t_\echo\simeq 2t_0$, where $t_0$ is the (instantaneous) pulse time.
The linear and chiral symmetric band structure is however not a necessary condition for this to happen. 
A weaker requirement is 
\be
\label{weakreq}
{\bf v}^+_\vk = -\xi_v{\bf v}^-_\vk, \quad \forall \vk,
\ee
or equivalently
\be
E^+_\vk = -\xi_v\, E^-_\vk + E_0,  \quad \forall \vk,
\label{eq:velrequirement}
\ee
where $E_0 \in \mathbb{R}$ is a constant energy offset and $\xi_v>0$ is a $\vk$-independent constant.
In this case every $\vk$-mode comes back to the origin at the same time $t_\echo$, because the distance traversed before the pulse, 
\be
 |{\bf d}_0^\vk| = |{\bf v}^\pm_\vk t_0|,
\ee
is the same as the distance $|\mathbf{d}_1^\vk|$ covered during time $ t_1 = \xi_vt_0$ after the pulse but in the opposite direction:
\begin{equation}
 \mathbf{d}_1^\vk = \mathbf{v}^\mp_\vk t_1 = -\frac{\mathbf{v}^\pm_\vk}{\xi_v}  \, \xi_vt_0 = -\mathbf{d}_0^\vk, \quad \forall \vk.
\end{equation}

Equation \eqref{eq:velrequirement} is for instance fulfilled in a direct band gap semi-conductor close to the extrema, 
where a quadratic expansion in $\vk$ of the band is justified.
If Eq.~\eqref{eq:velrequirement} does not hold exactly the time-reversal mechanism
is still effective, though it distorts the shape of the initial wave packet with different $\vk$-modes returning at different times. 
For the remainder of the paper, we assume two-band systems meeting Eq.~\eqref{eq:velrequirement}.

\subsection{Transition amplitudes and echo time}
\label{subsec:gentransamp}

The echo strength is determined by the efficiency of the population reversal induced by the homogeneous and diabatic pulse. 
For convenience we assume a pulse duration of the form
\be
 f(t) = \left\{\begin{array}{ll} 1 \,, \quad & t_0<t<t_0+\Delta t, \\ 
	0 \,, & \text{otherwise}.
\end{array}
\right.
\label{eq:f(t)}
\ee
The pulse is switched on and off instantaneously at times $t_0$ and $t_0 + \Delta t$ respectively,
while staying constant for its duration $\Delta t$.
As discussed in Ref.~\cite{reck2017}, the precise pulse shape is not critical for the QTM as long as it is diabatically activated/deactivated.  
The form \eqref{eq:f(t)} is thus chosen in order to simplify the analytical calculations. 

We consider the following general two-band Hamiltonian 
\begin{equation}
\label{fullH}
 H = H_0 + f(t)  H_1,
\end{equation}
where 
\begin{align}
 H_0&= h^0_0(\vk)\mathbbm{1} + \mathbf{h}_0(\vk)\cdot \vsigma , \label{eq:H0-gen} \\
 H_1(t)& =h^0_1(\vk)\mathbbm{1} + \mathbf{h}_1(\vk)\cdot \vsigma , \label{eq:H1-gen}
\end{align}
with $\vsigma = (\sigma_x,\sigma_y,\sigma_z)^T$ and $\mathbf{h}_\alpha = (h^x_\alpha,h^y_\alpha,h^z_\alpha )^T,\;\alpha=0,1$.
Unless otherwise specified we work in Fourier space, so that $\vk$ is a wave vector and not an operator, 
while $h^i_\alpha$ are real-valued functions of $\vk$, whose argument will be omitted in the following for
the sake of shortness.
The eigenstates of ${H}_0$ are normalized and complex-valued 2-spinors denoted by $|\varphi_{\vk}^{\pm} \rangle$, 
whose associated wave functions read
\be
\label{eigenstates}
\langle{\bf r}|\varphi_\vk^\pm\rangle = 
e^{i\vk\cdot{\bf r}} 
\left(
\begin{array}{c}
c^\uparrow_{\vk\,\pm} \\
c^\downarrow_{\vk\,\pm}
\end{array}
\right),\quad
|c^\uparrow_{\vk\,\pm}|^2 + |c^\downarrow_{\vk\,\pm}|^2=1.
\ee
Their energies are  
\begin{align}
 E^\pm_\vk = h^0_0 \pm \sqrt{(h^x_0)^2+(h^y_0)^2+ (h^z_0)^2}.
\end{align}
The eigenstates of the full Hamiltonian $H$ during the pulse are also 2-spinors of the form \eqref{eigenstates}.
They are indicated as $|\chi_{\vk}^{\pm} \rangle$, and the corresponding eigenenergies are
\begin{equation}
\varepsilon^\pm_\vk = h^0_0+h^0_1 \pm \hbar \Omega_\vk,
\end{equation}
with frequency
\begin{equation}
\Omega_\vk  = \frac{1}{\hbar} |{\bh}_0 + {\bh}_1| = \frac{1}{\hbar} \sqrt{\sum\limits_{i\in\lbrace x,y,z\rbrace}(h^i_0+h^i_1)^2}.
\end{equation}
Explicit expressions for $|\varphi_\vk^\pm\rangle,\;|\chi_\vk^\pm\rangle$ and the eigenenergies 
will be given for various exemplary cases in the following sections.

During the pulse the transition amplitude $|\varphi_\vk^{s} \rangle\,\rightarrow\,|\varphi_\vk^{-s} \rangle$ reads
\be
A_\vk =  \langle \varphi_\vk^\mp \mid U(t_0+\Delta t,t_0 ) \mid \varphi_\vk^\pm \rangle
\ee
with the time evolution operator
\begin{equation}
U(t_0+\Delta t, t_0 ) = \exp\left(-\frac{\iu}{\hbar} \Delta t   ({H}_0 + {H}_1)  \right).
\label{eq:timeevo0}
\end{equation}
In terms of the Hamiltonian (\ref{fullH}-\ref{eq:H1-gen}), it can be rewritten as
\begin{align}
U&(t_0+\Delta t, t_0) = \exp\left(-\frac{\iu}{\hbar} \Delta t  (h^0_0 + h^0_1)  \right)
\nn \\
&\times \left(\cos(\Omega_\vk \Delta t) \mathbbm{1} -\iu \frac{(\mathbf{h}_0 + \mathbf{h}_1) \cdot  \vsigma }{ \hbar\Omega_\vk  } \sin(\Omega_\vk  \Delta t) \right).
\label{eq:timeevo1}
\end{align}
Here, we used that for a unit vector ${\bf n}$
\be
\exp(i \alpha {\bf n}\cdot \vsigma) = \cos\alpha + i {\bf n}\cdot \vsigma \sin\alpha.
\ee
Orthonormality of the $|\varphi_\vk^\pm\rangle$ spinors thus yields the transition amplitude
\be
A_\vk = \me^{-\frac{i}{\hbar} \Delta t  (h^0_0 + h^0_1)}\frac{-i}{ \hbar\Omega_\vk} \sin(\Omega_\vk \Delta t) \, 
\langle \varphi_\vk^{\mp} \mid {\bf h}_1 \cdot  \vsigma \mid \varphi_\vk^\pm \rangle.
\ee
Splitting ${\bh}_1$ into its components parallel $(\parallel)$ and orthogonal $(\perp)$ to ${\bh}_0$,
\be
\mathbf{h}_1 = {\bh}_1^\parallel + {\bh}_1^\perp
\ee
with 
\be
\mathbf{h}^\perp_1\cdot \vsigma \mid \varphi_\vk^\pm \rangle = h_1^\perp \mid \varphi_\vk^\mp \rangle,
\ee
the transition amplitude further simplifies to
\be
A_\vk = -i\frac{h_1^\perp}{ \hbar\Omega_\vk } \sin\left(\Omega_\vk \Delta t\right) \me^{-\frac{i}{\hbar} \Delta t  (h^0_0 + h^0_1)},
\label{eq:gen-transamp}
\ee
which is one basic result we will frequently refer to in this work.
In this form, the transition amplitude $A_\vk$ of Eq.~\eqref{eq:gen-transamp} can be used, as in Ref.~\cite{reck2017}, to determine the echo strength for an arbitrary wave packet.
For each $\vk$ mode building such a wave packet, the velocity inversion probability -- and thus the echo strength -- 
is proportional to $|A_\vk|^2$.  Focusing for the moment on a single mode, 
the phase (\ie the exponential) is thus irrelevant for determining the QTM's effectiveness.  
The latter depends on tuning the pulse duration $\Delta t$
so as to maximize $\sin\left(\Omega_\vk \Delta t\right)$.  The most important term is however the prefactor
$|h_1^\perp| /(\hbar\Omega_\vk) \le 1$ with frequency 
\be
 \hbar\Omega_\vk = \sqrt{(h_1^\perp)^2+(h_0+h_1^\parallel)^2}.
\label{eq:gen-freq}
\ee
This is easiest to see by choosing without loss of generality a pseudospin basis 
such that $ {\bh}_0 =(0,0,h_0)^T$ and $ {\bh}_1 =(0,h_1^\perp,h_1^\parallel)^T$.
Thus, the magnitude of the first prefactor in Eq.~\eqref{eq:gen-transamp} becomes
\be
 \frac{|h_1^\perp|}{\hbar\Omega_\vk} = \dfrac{1}{\sqrt{1+ \left(\dfrac{h_0+h_1^\parallel}{h_1^\perp}\right)^2}},
 \label{eq:h1perp-rules}
\ee
which is largest (close to one) if $|h_1^\perp| \gg |h_0+h_1^\parallel|$.
Essentially, Eq.~\eqref{eq:h1perp-rules} signifies that the orthogonal component 
of the pulse ${\bh_1^\perp}$ should be the dominant energy scale of the problem, 
since it is the term that induces transitions between electron- and hole-like branches of the spectrum.
The results of the discussion thus far are summarized in Table \ref{table_reqs}.
\renewcommand{\arraystretch}{1.8} 
\begin{table}[h!]
\begin{tabular}{c|p{5.5cm}}
 Requirement for & \\
\hline\hline
 $H_0$ & \reqA ~$\nabla_\vk E^+_\vk = - \zeta_v \nabla_\vk E^-_\vk$ \\ \hline
 $H_1$ & \reqB ~$ {\bh}_1 \simeq {\bh}_1^\perp$, \ie  $h_1^\parallel \ll h_1^\perp $  \\ \hline
 \multirow{2}{2.5cm}{wave packet $\psi_0$} & \reqC ~$|h_1^\perp| \gg  |h_0(\vk)|, \quad \forall \vk$ relevant in $\psi_0$
\\ \cline{2-2}
& \reqD ~$|\sin(\Omega_\vk \Delta t)| \approx 1, \quad \forall \vk$ relevant in $\psi_0$
\end{tabular}
\caption{
Summary of requirements for an effective QTM.  Requirements \reqA~and \reqB~concern the general form
of the Hamiltonian, including the external driving.  They are necessary, though alone do not guarantee echoes with high fidelities.  
For the latter the additional conditions \reqC~and \reqD~need to be fulfilled, which also involve the form of the wave packet 
to be effectively time-mirrored.
}
\label{table_reqs}
\end{table}
\renewcommand{\arraystretch}{1}

We conclude this part by discussing further details of Eq.~\eqref{eq:gen-transamp}.
First, the phase factor $\me^{-\frac{i}{\hbar} \Delta t  (h^0_0 + h^0_1)}$ 
could lead to dephasing of a wave packet consisting of different $\vk$-modes, 
since here $h^0_0$ and $h^0_1$ are in general functions of $\vk$. 
However, if the velocity requirement \eqref{eq:velrequirement} or equivalently \reqA ~in Tab.~\ref{table_reqs} is met 
and $\partial (h_0^0+h_1^0)/\partial k = \xi_h \partial E_\vk^\pm /\partial k$, 
where $\xi_h\in \mathbb{R}$ is a $\vk$-independent constant, 
the phase will only cause a slight adjustment of the echo time $t_\echo$.  We will see this explicitly in the examples below. 
These conditions and the expression
\begin{equation}
 t_\echo = \left(1 + \xi_v\right) t_0 + \left( 1+  \xi_h \xi_v \right) \Delta t
\label{eq:gen_t-echo}
\end{equation}
for the general echo time are derived in App.~\ref{app:t_echo}.

Second, the echo time generally depends on the geometries of the two bands and of the initial wave packet as well. 
In particular, if the latter consists of a mixture of electron- and hole-like states, two echoes will arise if the energy branches
are not symmetric, \ie when one is steeper [``faster'', see Eq.~\eqref{weakreq}] than the other. 
States which have switched from the ``slow (fast)'' to the ``fast (slow)'' branch will cause the first (second) echo.

Third, the term $\sin\left(\Omega_\vk\Delta t\right)$ is responsible for a subtle effect, namely a splitting of the reflected wave packet
in the long-pulse duration limit.  We discuss this in detail in Sec.~\ref{subsec:pristine}.

%%%%%%%%%%% EXAMPLES %%%%%%%%%%%%%%%%%%%%%%%%%%%%%%%%%%%%%%%%%%%%%%%%%%%%%%%%%%%%

\section{QTM\lowercase{s} for various classes of band dispersions}
\label{sec:twoband-examples}

We put to the test the general formulas from Sec.~\ref{sec:genmod}
by considering systems with linear, hyperbolic and parabolic bands. 
For numerical reasons, we only consider one- and two-dimensional systems in this section, although the general results above are also valid in three dimensions.

\subsection{QTM for ideal Dirac bands}
\label{subsec:pristine}
First, let us consider the Dirac QTM put forward in Ref.~\cite{reck2017},
and extend its discussion to the yet unexplored regime of long pulses.  By this we mean pulses 
whose duration $\Delta t$ is long compared to certain internal time scales of a propagating wave packet
-- to be defined below -- though their switch on and off procedures stay diabatic. 
\subsubsection{Preliminaries}
Take a Dirac Hamiltonian with a pulsed mass gap \cite{reck2017}
\begin{equation}
\label{eq:Dirac}
 H =  \hbar v_F \vk \cdot \vsigma + f(t) M \sigma_z = H_0 + f(t) H_1.
\end{equation}
The time-dependence is as in Eq.~\eqref{eq:f(t)} and $\vk = (k_x, k_y,0)$.  
The eigenenergies are
\ber
& H_0\;\rightarrow\quad & E^\pm_\vk =  \pm \hbar v_F k,  \label{eq:E-Dirac}
\\
& H\;\,\rightarrow\quad & \varepsilon^\pm_\vk = \pm M \sqrt{1+\kappa^2},
\eer
with $\kappa = E^+_\vk/M$. 
Furthermore $h_0^0 = h_1^0 = 0$ implies
\begin{equation}
 \hbar \Omega_\vk = \varepsilon^+_\vk.
\end{equation}
The $H_0$ eigenvectors read
\be
 |\varphi^\pm_\vk\rangle = \frac{1}{\sqrt{2}} \begin{pmatrix}
                                               1 \\ \pm \me^{i\theta_\vk}
                                              \end{pmatrix} 
|\vk\rangle,
\label{eq:eigvec-pristgraph} 
\ee
with $\theta_\vk$ the polar angle of $\vk$.
Thus, the pulse is such that 
${\bh}_1 \perp {\bh}_0$, \ie ${\bh}_1 = {\bh}_1^\perp$, and meets condition \reqB~in Sec.~\ref{subsec:gentransamp}.
The transition amplitude \eqref{eq:gen-transamp} becomes 
\begin{equation}
 A_\vk =  -\frac{\iu}{\sqrt{1+\kappa^2}} \sin \left(\frac{M \Delta t }{\hbar} \sqrt{1+\kappa^2}\right),
\label{eq:transampl-pristgraph}
\end{equation}
having used that $h_1^\perp = M$.
Condition \reqC ~from Sec.~\ref{subsec:gentransamp} yields
\begin{equation}
 |h_1|^\perp \gg |\mathbf{h}_0| \quad \Leftrightarrow \quad M \gg E^+_\vk \quad \Leftrightarrow \quad \kappa \ll 1.
\end{equation}
Since the bands are symmetric ($\xi_v$\;=\;$1$) and lack pseudospin-independent terms ($\xi_h$\;=\;$0$)
the echo time is
\begin{equation}
 t_\echo = 2t_0 + \Delta t.
\end{equation}

\subsubsection{Long pulse durations}
\label{subsubsec:verylongdt}

%%%%%%%%%%%%%%%%%%%%%%%%%%%%% FIGURE %%%%%%%%%%%%%%%%%%%%%%%%%%%%%%%%%%%%%%%%%%%%%
\begin{figure*}
 \centering 
%     \def\svgwidth{\textwidth}
%     \input{./fig2.pdf_tex}
% % 
     \includegraphics[width=\textwidth]{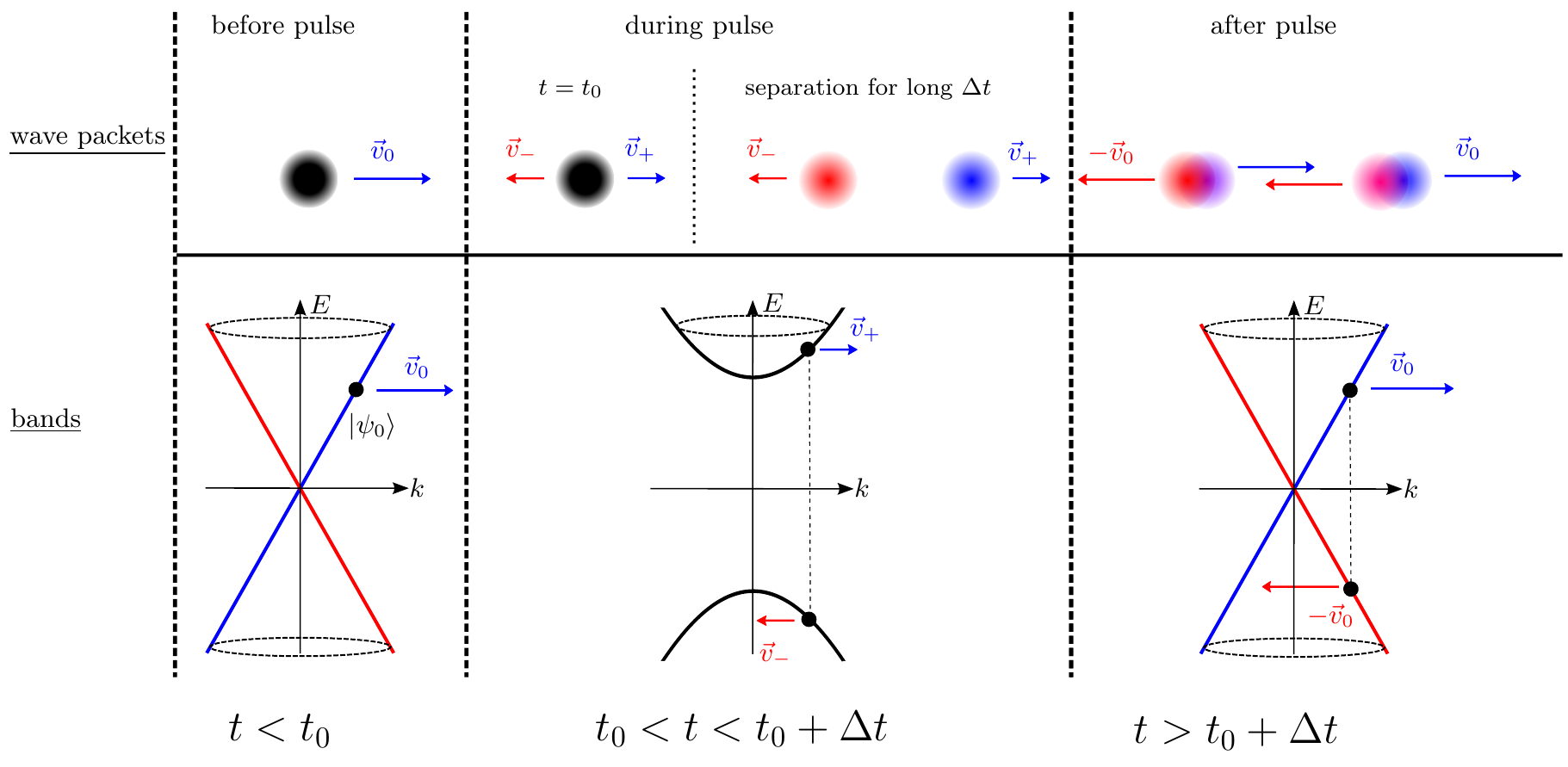}
\caption{
Motion of the wave packet for such long pulse durations that it splits significantly during the pulse ($\Delta t |\mathbf{v}^+_\vk| \gg  1/\Delta k$). 
After switching on the pulse ($t>t_0$), the wave packet separates into two partial wave packets (mid panel), because its support lives in both bands which have opposite velocity directions. 
After switching off the pulse, each partial wave packet splits again. 
Since they are separated, no interference of the ''reflected states`` takes places, which was responsible for the strong $\Delta t$-dependence of the echo in the small $\Delta t$ case.
Instead two distinct echoes are expected.
}\label{fig:longdt-mechanism}
\end{figure*}
%%%%%%%%%%%%%%%%%%%%%%%%%%%%% FIGURE %%%%%%%%%%%%%%%%%%%%%%%%%%%%%%%%%%%%%%%%%%%%%

\begin{figure*}
 \centering 
     \includegraphics[width=\textwidth]{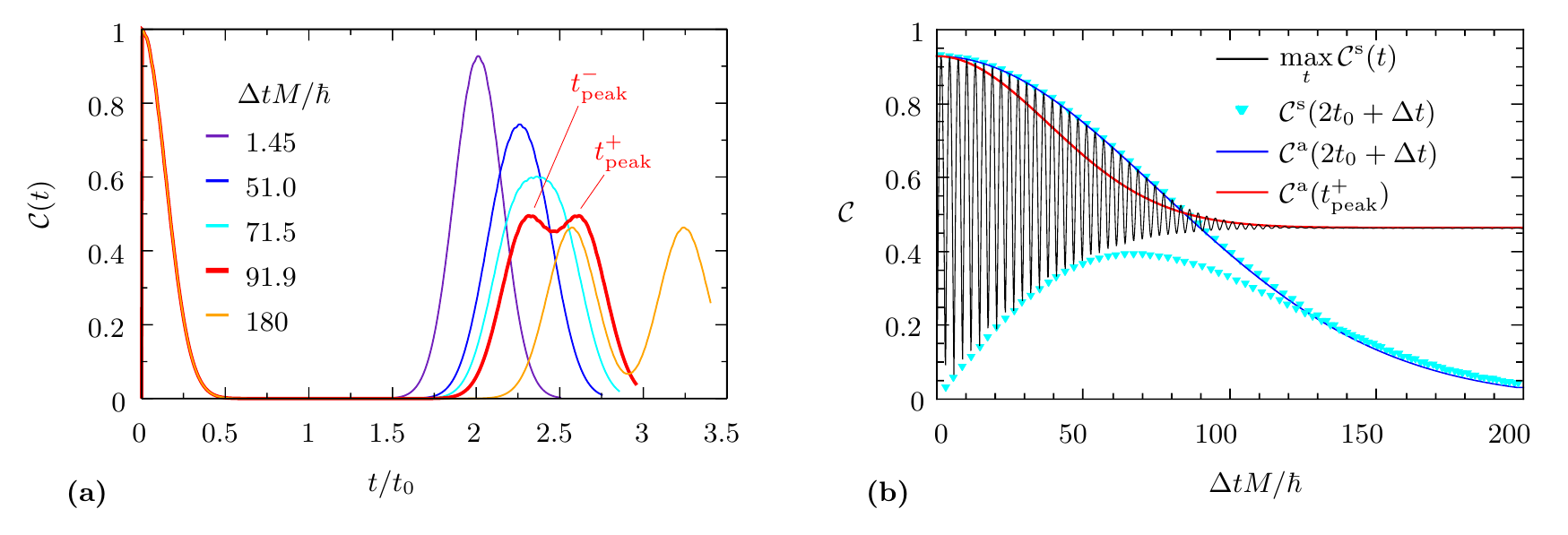} 
\caption{
    Echo in ideal Dirac bands with long pulse duration $\Delta t$.  ($t_0 = 200 \hbar/M$)
    (a) Correlation plotted for different values of $\Delta t$ such that the sine becomes one (only maximal values), \ie $M\Delta t /\hbar \sqrt{1+\kappa_0^2} = (2n+1) \pi/2$ with $n\in \mathbb{N}_0$. 
    For higher $\Delta t$, the echo (peak in correlation) appears later. 
    At around $\Delta t = 90 \hbar/M$ two peaks arise (red) due to the splitting of the wave packet during the pulse.
    Therefore two separate wave packets are going back to the initial position. 
    The splitting becomes more and more pronounced for even higher $\Delta t$. 
    (b) Peak values and the values at $t=2t_0+\Delta t$ of the correlation, where $\mathcal{C}^s$ and $\mathcal{C}^a$ denote the correlations obtained by simulation and analytics, respectively. 
    The black line shows results obtained by simulation, where the maximum value of the correlation for each simulation is extracted. 
    The oscillations vanish when the wave packets split significantly during the pulse. 
    The cyan triangles show again simulation results, but instead of taking the maximum, the correlation at $t=2t_0+\Delta t$ is plotted (only envelopes). 
    It fits very well the expected curve (dark blue, only maxima) of Eq.~\eqref{eq:Corr(prevechot)}. 
    Finally, the red line shows the analytical expected echo of Eq.~\eqref{eq:Corr(newechot)} at $t_\peak = 2t_0+\Delta t+t_2$, which matches well the simulation data for large $\Delta t$.
  }
\label{fig:verylong_dt}
  \end{figure*}

A sufficiently long $\Delta t$ reveals
the true nature of the factor $\sin\left(\Omega_\vk \Delta t\right)$ in the transition amplitude \eqref{eq:gen-transamp},
namely that of a superposition of two plane wave states $\langle{\bf r}|\chi_\vk^\pm\rangle$ belonging to the full
Hamiltonian \eqref{fullH}.  Such states, into which a given $H_0$ initial mode $\langle{\bf r}|\varphi_\vk^\pm\rangle$ 
is decomposed during the pulse, propagate with opposite velocity and eventually lead to a splitting of the initial wave packet,
see Fig.~\ref{fig:longdt-mechanism}.
For a short $\Delta t$ the two counter-propagating components of the wave packet still overlap and interfere, 
which yields the oscillatory behavior of the echo strength as a function of $\Delta t$, see Fig.~\ref{fig:verylong_dt}(b).
In the long $\Delta t$ limit however the counter-propagating components separate, coalescing into two separate
wave packets moving away from each other.
Their components $\langle{\bf r}|\chi_\vk^\pm\rangle$ are decomposed back into $H_0$ modes $\langle{\bf r}|\varphi_\vk^\pm\rangle$
when the pulse is switched off, so that two distinct time-mirrored wave packets move back to the origin
and result in two echo peaks, see red and yellow curves in Fig.~\ref{fig:verylong_dt}(a).
\footnote{Two other partial wave packets propagate instead in the initial forward direction and do not lead to any echo.} 
Quantitatively, given a wave packet with reciprocal space width $\Delta k$ centered around $\vk=\vk_0$, 
it is shown in App.~\ref{app:t_echo-longDeltat} that if $\Omega_\vk$ is smooth over the scale $\Delta k$ the distinct echo times are
\be
\label{eq:twotimes}
 t^\pm_\peak = 2t_0 +\Delta t \left(1\pm   \frac{\kappa_0}{\sqrt{1+\kappa_0^2}}\right),
\ee
where $\kappa_0 = |E^\pm_{\vk_0}|/M$.
The amplitude of each peak is exactly half the best achievable echo, 
obtained when the sine factor in Eq.~\eqref{eq:transampl-pristgraph} equals one.
Yet, if the two time-reversed wave packets still overlap, 
the strongest echo remains at $t_\echo = 2t_0 + \Delta t = (t^+_\peak+t^-_\peak)/2$.
On the contrary, for larger $\Delta t$ the overlap decreases and the two equivalently strong echo peaks are resolved.
In obtaining Eq.~\eqref{eq:twotimes} we also neglected the spreading of the wave packet during the pulse $\Delta t$,
which numerics show indeed to be irrelevant in the parameter range that we will consider.

%%%%%%%%%%%%%%%%%%%%%%%%% FIGURE %%%%%%%%%%%%%%%%%%%%%%%%%%%%%%%%%%%%%%%%%%%%%%%%%%%%%%%%%%%%%%%%%%
%%%%%%%%%%%%%%%%%%%%%%%%% FIGURE %%%%%%%%%%%%%%%%%%%%%%%%%%%%%%%%%%%%%%%%%%%%%%%%%%%%%%%%%%%%%%%%%%

Figure \ref{fig:verylong_dt} shows explicit results for an initially Gaussian wave packet of the form
\be
\label{eq:wavepacket}
 \psi_0(\vk) = \frac{1}{\sqrt{2\pi}\Delta k} \exp\left(-\frac{(\vk-\vk_0)^2}{2\Delta k^2} \right) \begin{pmatrix}
                                                                                                  1 \\ \me^{i\theta_\vk}
                                                                                                 \end{pmatrix},
\ee
whose real space magnitude is
\begin{equation}
 |\psi_0(\br)| = \frac{\Delta k}{\sqrt{\pi}} \exp\left(-\frac{|\br|^2 \Delta k^2}{2} \right).
\end{equation}
The time-evolution of the initial wave packet is performed using the ''Time-dependent quantum transport`` (TQT) package 
\cite{krueckl2013}. 
The time-evolution operator is expanded in Chebychev polynomials.
To avoid spurious fermion-doubling problems, which arise when discretizing linear dispersions, momentum (position) operators are applied to the state in momentum (real) space by (inverse) fast Fourier transformations of the wave function. 
TQT is employed for all simulations throughout this paper.

The relative energy parameter is chosen to be $\kappa_0$\;=\;$\hbar v_F |\vk_0|/M$\;=\;$0.4$.
The wave packet width is narrow in $\vk$-space and taken as $\Delta k = k_0/8$ here and in the rest of the paper.
Pulses are long (short) if $\Delta t \gg\,(\ll)\, \Delta x/2|{\bf v}^{\pm}_\vk|$
for all $\vk$ modes building up the wave packet of Eq.~\eqref{eq:wavepacket}, where $\Delta x= 1/\Delta k$ is the real space width of the Gaussian wave packet.  This condition ensures
the separation (overlap) of the counter-propagating sub-wave packets shown in the central panel of Fig.~\ref{fig:verylong_dt}.
As a measure of the echo strength we employ the density correlator \cite{reck2017}
\begin{equation}
\mathcal{C}(t) = \int \,{\rm d}^2\br \, |\psi(\br,0)| |\psi(\br,t)| \label{eq:spaceCorr}.
\end{equation}
The approximations which lead to Eq.~\eqref{eq:twotimes} now yield 
\begin{align}
 |\psi_\text{refl}(\br,t)| \approx \frac{\Delta k}{\sqrt{\pi}} \,\frac{1}{\sqrt{1+\kappa_0^2}} \, \Bigg| \sum\limits_{l=\pm1}\frac{l}{2} \me^{\iu l  \Delta t \varepsilon_+(k_0)/\hbar}  \nonumber\\
\times \exp\left(-\frac{(\br+ v_F \hat{\vk}_0(t_0-t^\prime-l t_2 ))^2 \Delta k^2}{2} \right)\Bigg|.
\label{eq:verylong-dt_psir-0}
\end{align}
for the density after the pulse, $t = t_0+\Delta t + t^\prime$, which has undergone the transition to the other band. 
The unit vector in the direction of $\vk_0$ is denoted $\hat{\vk}_0$, and a similar convention will be used throughout this work.
The term $\sin\left(\Omega_\vk\Delta t\right)$ was rewritten
in exponential form, and spreading during the pulse duration was neglected, as appropriate for narrow wave packets
$\Delta k\ll k_0$.

For the analytical estimation of the maximal echo strength, 
we consider only $ \Delta t\,\varepsilon^+_{k_0}/\hbar = (2n+1) \pi/2 $, 
with $n \in \mathbb{N}_0$, such that
\be
 l \me^{i l  \Delta t\,\varepsilon^+_{k_0}/\hbar} = i \sin[(2n+1)\pi/2] = \pm i
\ee
for both values $l=\pm 1$.
Furthermore, we limit ourselves to the 1D line along the propagation direction of the wave packet center of mass, 
which we choose as the $x$-axis:
\begin{align}
& \mathcal{C}(t) = \int \mathrm{d}x \, |\psi_0(x)| |\psi(x,t)| \nonumber \\
&= \frac{\Delta k}{\sqrt{\pi}} \,\frac{1}{2\sqrt{1+\kappa_0^2}} \sum\limits_{l=\pm1}  \int \mathrm{d}x \me^{-\frac{ \Delta k^2}{2} (x^2 + (x+ v_F (t_0-t^\prime-l t_2 ))^2)} \nonumber \\
&= \frac{1}{2\sqrt{1+\kappa_0^2}} \sum\limits_{l=\pm1}  \exp\left(-\frac{ \Delta k^2v_F^2 }{4} (t_0-t^\prime-l t_2 )^2\right).
\end{align}
At the symmetric echo time $t= 2t_0 +\Delta t$, \ie $t^\prime = t_0$, the correlator reads
\begin{equation}
\mathcal{C}(2t_0+\Delta t) =\frac{1}{\sqrt{1+\kappa_0^2}} \exp\left(-\frac{ \Delta k^2v_F^2 \Delta t^2}{4}  \, \frac{\kappa_0^2}{1+\kappa_0^2}\right).
\label{eq:Corr(prevechot)}
\end{equation}
At the return times of the two split wave packets, $t=t^\pm_\peak$, \ie $t^\prime = t_0 \pm t_2$, one has instead
\begin{equation}
\mathcal{C}(t^\pm_\peak) =\frac{1}{2\sqrt{1+\kappa_0^2}} \left( 1+ \exp\left(- \Delta k^2v_F^2 \Delta t^2  \, \frac{\kappa_0^2}{1+\kappa_0^2}\right)\right),
\label{eq:Corr(newechot)}
\end{equation}
which becomes larger than $\mathcal{C}(2t_0+\Delta t)$ in Eq.~\eqref{eq:Corr(prevechot)} for large enough $\Delta t$.

Figure \ref{fig:verylong_dt}(a) shows the numerically computed density correlator \eqref{eq:spaceCorr} for different pulse lengths.  
As expected, for small $\Delta t$ the echo is at $t_\echo = 2t_0 +\Delta t$, while for large $\Delta t$, \eg$\Delta t = 90 \hbar/M$ 
the splitting of the mirrored wave packet is obvious, and increases for larger $\Delta t$.

We plot single correlator values (``echo strength'') for each simulation as a function of $\Delta t$ in Fig.~\ref{fig:verylong_dt}(b). 
The black line and cyan triangles respectively show the simulated data with TQT of the maximal value of $\mathcal{C}(t)$, 
and $\mathcal{C}(2t_0+\Delta t)$ (only minimal and maximal envelopes to avoid overcrowding the figure). 
For small $\Delta t$ they match perfectly, but at $\Delta t \gtrsim 60 \hbar/M$ deviations show up. 
The blue and red lines show the corresponding analytical estimations of $\mathcal{C}(2t_0+\Delta t)$, Eq.~\eqref{eq:Corr(prevechot)},
and $\mathcal{C}(t^\pm_\peak)$, Eq.~\eqref{eq:Corr(newechot)}, respectively.  As expected from the previous discussion, the former catches the short time
behavior ($\Delta t \ll \Delta k/2|{\bf v}^{\pm}_\vk|$),
the latter the long time one, when the initial wave packet has split
($\Delta t \gg \Delta k/2|{\bf v}^{\pm}_\vk|$).  
In both cases the agreement with the numerics is excellent.

These results can be of experimental interest, since for long $\Delta t$ the QTM is not affected by pulse length fluctuations,
as opposed to the small-$\Delta t$ regime.
Furthermore, the long-$\Delta t$ behavior is universal and applies to any band structure qualifying for time-mirroring,
in particular those to be studied in Secs.~\ref{subsec:AsymLin}, \ref{subsec:hyp-bands} and \ref{subsec:parabol-bands}.
However, in order not to overburden the following discussions, from now on we will focus on short pulses only.

%%%%%%%%%%%%%%%%%%%%%%%%%%%%%%%%   Asymmetric linear bands   %%%%%%%%%%%%%%%%%%%%%%%%%%%%%%%%%%

\subsection{Asymmetric linear bands}
\label{subsec:AsymLin}

%%%%%%%%%%%%%%%%%%%%%%%%% FIGURE %%%%%%%%%%%%%%%%%%%%%%%%%%%%%%%%%%%%%%%%%%%%%%%%%%%%%%%%%%%%%%%%%%5
\begin{figure*}
 \begin{center}
     \includegraphics[width=0.8\textwidth]{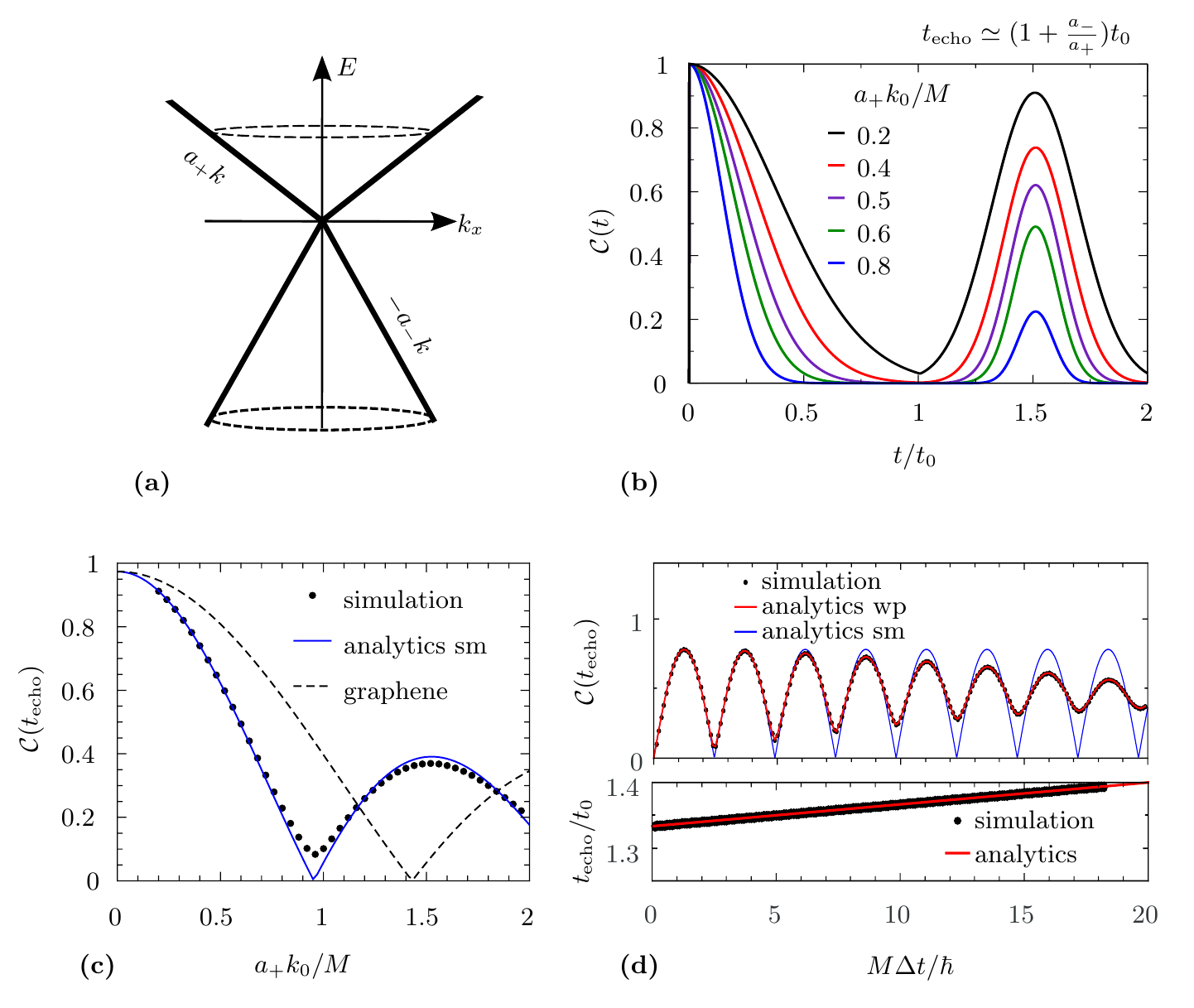} 
 \end{center}
\caption{
(a) Example of the band structure of the Hamiltonian in Eq.~\eqref{eq:asym_Dirac} for $a^+ > a^-$. 
(b) The correlation of Eq.~\eqref{eq:spaceCorr} for different wave packet energies starting in the positive energy branch for an arbitrarily chosen pulse length such that $M\Delta t/\hbar = 1.8$. 
The Hamiltonian parameters are $a^- = 2 a^+$, which is why the velocity after the transition is twice as large and the echo takes place already at $t=1.5 t_0$. 
The different decay rates of the correlation are due to different wave packet widths given by the relation $\Delta k  =  k_0/8$, where $\Delta k$ is the width of wave packet in reciprocal space. 
(c) Echo strength as a function of the initial energy $a^+ k_0$ relative to the pulse strength $M$. The simulation (black dots) is compared to the modulus of the transition amplitude of Eq.~\eqref{eq:linasym-transamp} (blue). 
The deviations are due to the finite width of the wave packet as above in the graphene case. 
The black dashed line shows analytically the case of symmetric bands, \ie $a^-=a^+$.
(d) Dependence of echo time (below) and echo strength (above) on the pulse duration $\Delta t$ for fixed $M$, both matching very well the expectation.
}  
\label{fig:asym-linear}
\end{figure*}
%%%%%%%%%%%%%%%%%%%%%%%%% FIGURE %%%%%%%%%%%%%%%%%%%%%%%%%%%%%%%%%%%%%%%%%%%%%%%%%%%%%%%%%%%%%%%%%%5

Let us look at the QTM in a linear but chiral asymmetric band structure, as shown in Fig.~\ref{fig:asym-linear}(a). 
This is meant to be a toy model to show the effect of different velocities in the two bands.

The Hamiltonian reads
\ber
H_0 
&=&  
\frac{a^+-a^-}{2} |\vk| \mathbbm{1} + \frac{a^+ + a^-}{2} \vk\cdot \vsigma 
\nn \\
&=& 
\begin{pmatrix} 
\frac{a^+-a^-}{2} |\vk| & \frac{a^+ + a^-}{2}(k_x-\iu  k_y)  \\ \frac{a^+ + a^-}{2}(k_x+\iu  k_y)  & \frac{a^+-a^-}{2} |\vk|        
\end{pmatrix},
\label{eq:asym_Dirac}  
\eer
with $a^\pm>0$ and eigenenergies
\begin{align}
 E^\pm_\vk &= \left(\frac{a^+-a^-}{2} \pm \frac{a^+ + a^-}{2}\right) |\vk|, \nonumber \\
 E^+_\vk &= a^+ |\vk|, \nonumber \\ 
 E^-_\vk &= -a^- |\vk|.
\end{align} 
Since the structure of the Hamiltonian is the same as in Eq.~\eqref{eq:Dirac}, 
except for the additional $\vk$-dependent term on the diagonal ($\propto \mathbbm{1}_{2\times2}$),
the eigenvectors $|\varphi_\pm^\vk\rangle$ are still given by Eq.~\eqref{eq:eigvec-pristgraph}.
The transition operator is again $\sigma_z$, independently of $\vk$,  
and the pulse is taken as
\begin{equation}
 {H}_1 = M \sigma_z.
\end{equation} 
With $M\gg a^\pm|\vk_0|$ and a narrow wave packet peaked at $\vk_0$,
the conditions \reqA, \reqB~and \reqC~in Tab.~\ref{table_reqs} are fulfilled.
The Rabi-type frequency differs slightly from the chiral symmetric case and reads
\begin{equation}
 \hbar \Omega^\vk = M \sqrt{1+ \tilde\kappa^2},
\end{equation}
where $\tilde\kappa = \tfrac{a^+ + a^-}{2} k/M$. 
Furthermore, the transition amplitude acquires an additional phase from to the terms $\propto\mathbbm{1}$
in Eq.\eqref{eq:asym_Dirac},
\begin{equation}
 A_\vk = -\iu\frac{1}{\sqrt{1+ \tilde\kappa^2}} \sin\left(\frac{M \Delta t}{\hbar}\sqrt{1+ \tilde\kappa^2} \right) \me^{-\frac{\iu}{\hbar} \Delta t  (a^+-a^-)k} .
\label{eq:linasym-transamp}
\end{equation}
As discussed in Sec.~\ref{subsec:gentransamp} and App.~\ref{app:t_echo}, this additional phase shifts the echo time.
The latter now reads
\begin{equation}
 t_\echo =\left(1+ \frac{a^s}{a^{-s}}\right) t_0 + \left( 1 +  \frac{a^+ - a^-}{2a^{-s}}\right) \Delta t + \mathcal{O}(\Delta k \Delta t),
\label{eq:echotime-}
\end{equation}
where $s=\pm$ denotes the band occupied by the initial wave packet.

The characteristics of the QTM for asymmetric Dirac bands are shown in Fig.~\ref{fig:asym-linear}, panels (b), (c) and (d).   
The initial wave packet is launched without loss of generality into the electron-like band ($s=+$).
The most important difference compared to the chiral symmetric case is clearly visible in Fig.~\ref{fig:asym-linear}(b): 
the fraction $a^-/a^+$ determines the echo time.  Here $a^-/a^+ = 2$, meaning that the magnitude of the initial wave packet velocity 
is half that after the pulse, and thus $t_\echo \simeq 1.5 t_0$ rather than $2t_0$. 
The initial energy $E^+_{\vk_0}$ is varied between $0.2M$ and $0.8M$, causing corresponding variations in the echo strength.
The wave packet width in $\vk$-space is $\Delta k = k_0/8$.
The echo strength $\mathcal{C}(t_\echo)$ is shown in panel (c) as a function of the initial 
wave packet energy $E^+_{\vk_0}=a^+k_0$ (with $\Delta t = 1.8 \hbar/M$, $a^- = 2a^+$), and in the upper half of panel (d)
as a function of $\Delta t$ ($E^+_{\vk_0} = 0.4 M$, $a^- = 3a^+$). 
Here and in the rest of the paper, ``analytics sm'' indicates analytical results valid for a {\bf s}ingle {\bf m}ode, 
and means that the echo strength is approximated by the modulus of the transition amplitude at the mean wave vector $\vk_0$,
$A_{\vk_0}$.  On the other hand, ``analytics wp'' indicates results that take into account 
the transition amplitudes $A_\vk$ of the different modes in the {\bf w}ave {\bf p}acket, following Ref.~\cite{reck2017}.

For comparison, the black dashed line in panel (c) indicates the 
(analytical) echo strength for the chiral symmetric Dirac setup, 
\ie $a^- = a^+$.
The lower half of panel (d) shows the echo time $t_\echo$ as a function of the pulse duration.
Analytics and numerics are in excellent overall agreement.
In particular, the slight increase of $t_\echo$ as a function of $\Delta t$ matches 
perfectly the expected dependence given by Eq.~\eqref{eq:echotime-},
\begin{equation}
 t_\echo = \frac{4}{3} t_0 + \frac{2}{3} \Delta t.
\end{equation}

This simple example shows explicitly that in linear bands the magnitude of the electron- and hole-like velocities do not have to be equal
for the QTM to work effectively.  The only asymmetry-induced effect is a change in $t_\echo$.
In the following we will perform a similar analysis of the QTM
by furthermore allowing different velocities for each $\vk$ mode -- a result of considering 
non-linear band dispersions.

%%%%%%%%%%%%%%%% Hyperbolic bands %%%%%%%%%%%%%%%%%%%%%%%%%%%%%%%%%%%%%%%%%%%%%%%%%%%%%%%%%%%%%%%%%%%%%%%%%%%%

\subsection{Hyperbolic bands }
\label{subsec:hyp-bands}
%%%%%%%%%%%%%%%%%%%%%%%%% FIGURE %%%%%%%%%%%%%%%%%%%%%%%%%%%%%%%%%%%%%%%%%%%%%%%%%%%%%%%%%%%%%%%%%%5
\begin{figure*} 
 \begin{center} 
     \includegraphics[width=\textwidth]{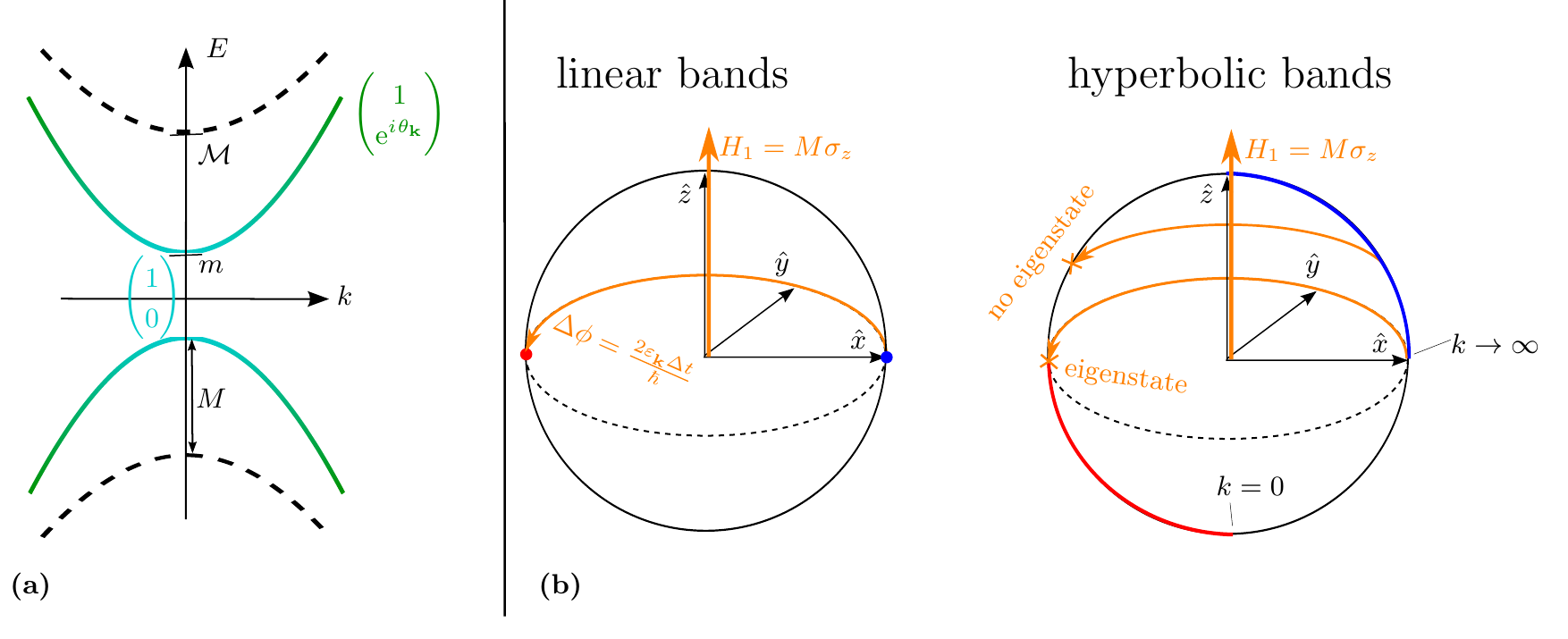} 
 \end{center}
\caption{(a) Hyperbolic band structure of the Hamiltonian $H_0$ in Eq.~\eqref{eq:hyperbol-Hamil} with the eigenstates pseudospin structure color-coded. 
(b) Visualization of the transitions on the Bloch sphere which are induced by the mass gap pulse in both linear and hyperbolic band structures.
For simplicity, we choose a certain direction of $\vk$, here $\vk = (k,0,0)$. 
Parts of the Bloch sphere highlighted in blue (red) are those visited by the positive (negative) helicity eigenspinor 
while varying $k\in\mathbb{R}_+$.  For linear bands the eigenspinor is located in the equatorial plane $\forall\,\vk$, 
which implies that changing $k$ does not change its direction (single points on the Bloch sphere).   
This is not the case for the hyperbolic bands, where varying $k$ leads to a change of the eigenspinor direction 
(meridian arcs on the Bloch sphere), as shown in panel (a). 
The pulse Hamiltonian $H_1=M\sigma_z$ (orange vertical arrow) leads to a rotation of the states in the Bloch sphere (orange curved arrow), analogous to a magnetic field for spins. 
While for linear bands the eigenstate can be mapped to the according eigenstate with negative energy for any $k$ (see \reqA), this is only possible for large $k$ in the hyperbolic case (lower curved orange arrow). For small $k$ compared to the initial gap $M_0$, the pulse Hamiltonian does not lead to a perfect mapping to the opposed eigenstate (upper curved orange arrow), but to a superposition, which diminishes the echo, compare Fig.~\ref{fig:hyperbolic-data}(a).
}  
\label{fig:hyperbolic}
\end{figure*}
%%%%%%%%%%%%%%%%%%%%%%%%% FIGURE %%%%%%%%%%%%%%%%%%%%%%%%%%%%%%%%%%%%%%%%%%%%%%%%%%%%%%%%%%%%%%%%%%5

\subsubsection{Mass gap pulse}
\label{subsubsec:mass_gap}
Take a hyperbolic band structure described by
\begin{equation}
 H = \hbar v_F \vk\cdot \vsigma + m \sigma_z + f(t)\, M  \sigma_z = H_0 + f(t) H_1
\label{eq:hyperbol-Hamil}
\end{equation}
with constant and time-dependent components of the mass gap, denoted respectively $m$ and $M$.
The eigenenergies of $H_0$ are 
\be
\label{eq:hyperenergies}
 E^\pm_\vk = \pm m\sqrt{1+\kappa_m^2}
\ee
with $\kappa_m = \hbar v_F k/m$.  The system is chiral symmetric,
the velocity fulfills ${\bf v}^\pm_\vk=-{\bf v}^\mp_\vk\;\forall\,\vk$
and thus the requirement of Sec.~\ref{subsec:genvelreq} with $\xi_v=1$.
Due to the lack of diagonal terms in $H$ we have $\xi_h = 0$, 
and therefore expect the echo to take place at $t_\echo = 2t_0+\Delta t \simeq 2t_0$ as in the (short pulse) symmetric Dirac case.
This holds irrespective of a qualitative difference with the latter scenario: 
The wave packet now spreads due to different $\vk$-mode velocities. 
Such a spreading is however undone by the QTM, since both slow and fast modes make (vertical) transitions to the
corresponding velocity-inverted slow and fast modes: The latter, which left the former behind in the ``forward'' 
time-evolution before the pulse, will catch up with them again on the ``backward'' time-evolution after the QTM has acted.  
Spreading has therefore no effect on the echo strength.

%%%%%%%%%%%%%%%%%%%%%%%%% FIGURE  sig_x pulse  %%%%%%%%%%%%%%%%%%%%%%%%%%%%%%%%%%%%%%%%%%%%%%%%%%%%%%%%%%%%%%%%%%5
\begin{figure*}
 \centering 
     \includegraphics[width=0.8\textwidth]{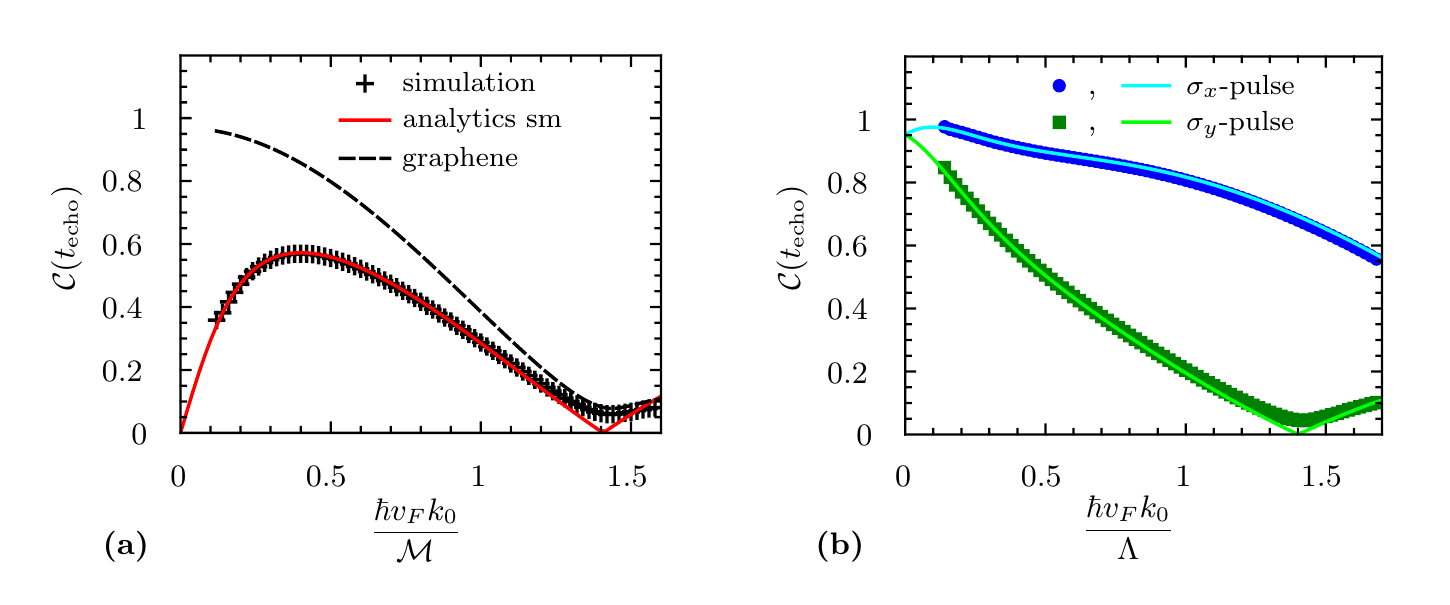} 
 \centering 
\caption{
(a) Echo strength as a function of the ratio between mean wave number and total gap, $\hbar v_F k_0 /\mathcal{M} $, with a pulse duration $\Delta t=  1.8 \hbar /M$. The simulation (black crosses) is compared to the modulus of the transition amplitude of Eq.~\eqref{eq:transamp-hyper} for the single mode $\vk_0$ (red). 
% The deviations are due to the finite width of the wave packet as above in the graphene case. The black dashed line shows the corresponding results of the simulation for the graphene case, \ie symmetric linear bands. 
(b) Comparison of the simulated data for the echo strength with the analytical expected strength given by the transition amplitude (again single mode) in Eq.~\eqref{eq:transamp-hyperbol_sigx} as a function of the ratio of the mean wave number of the wave packet and the pulse amplitude $\hbar v_F k_0 /M_1$, which shows a good agreement in both cases.
}  \label{fig:hyperbolic-data}
\end{figure*}
%%%%%%%%%%%%%%%%%%%%%%%%% FIGURE %%%%%%%%%%%%%%%%%%%%%%%%%%%%%%%%%%%%%%%%%%%%%%%%%%%%%%%%%%%%%%%%%%5

There is however a further difference with linear bands that does matter for the echo strength, 
namely the different $H_0$ eigenstates' structure:
\be
\label{eq:hyperspinors}
|\varphi^\pm_\vk\rangle = \frac{1}{\sqrt{2}\sqrt{1+\kappa_m^2\pm\sqrt{1+\kappa_m^2}}} 
\begin{pmatrix}
1\pm\sqrt{1+\kappa_m^2} \\ \kappa_m \,\me^{\iu \theta_\vk}                                         
\end{pmatrix}
|\vk\rangle.
\ee
In massless Dirac dispersions, $\sigma_z$ represents the 
$|\varphi^\pm_\vk\rangle\leftrightarrow|\varphi^\mp_\vk\rangle$ transition operator independently of $\vk$,
while now this holds only for $\kappa_m \gg 1$, see Fig.~\ref{fig:hyperbolic}(a).
As shown in Fig.~\ref{fig:hyperbolic} (b), for smaller wave vectors the spinors of Eq.~\eqref{eq:hyperspinors} 
progressively align with the $z$-direction, and for $\kappa_m = 0$
they coincide with the $\sigma^z$ eigenspinors $(1,0)$ and $(0,1)$ as evident from Eq.~\eqref{eq:hyperbol-Hamil}.
At $\kappa_m=0$ the transition operator is then a linear combination of $\sigma_x$ and $\sigma_y$. 
As a consequence, depending on the initial wave packet energy,
different pulse Hamiltonians $H_1$ satisfy condition \reqB~from Sec.~\ref{subsec:gentransamp}. 
The $\sigma_z$ pulse appearing in Eq.~\eqref{eq:hyperbol-Hamil} will thus yield good echoes
only away from the $\kappa_m=0$ point.

Denoting the total gap during the pulse $\Mcal=m + M$, and $\kappa = \hbar v_F k/\Mcal$, one has
\ber
\hbar \Omega_\vk &=& \Mcal \sqrt{1+\kappa^2} 
\\
h_1^\perp &=& M \langle \varphi_\vk^{-\mp} \mid \sigma_z \mid\varphi_\vk^\pm \rangle = -M \frac{\kappa_m}{\sqrt{1+\kappa_m^2}} 
\nn \\
	  &=& -M \frac{\kappa}{\sqrt{\frac{m^2}{\Mcal^2}+\kappa^2} },
\eer
which leads to the transition amplitude
\be
A_\vk = i \frac{\kappa}{\sqrt{\dfrac{m^2}{\Mcal^2}+\kappa^2}} \frac{M}{\Mcal}\frac{1}{\sqrt{1+\kappa^2}} \sin\left( \frac{\Mcal\Delta t}{\hbar} \sqrt{1+\kappa^2} \right).
\label{eq:transamp-hyper}
\ee

Fig.~\ref{fig:hyperbolic-data}(a) shows the echo strength of a Gaussian wave packet with a $\vk$-space width of $\Delta k = k_0/8$,
plotted as a function of $\kappa$.  The approximate analytical result
$|A_{\vk_0}|$ (red), with the transition amplitude \eqref{eq:transamp-hyper}, is compared with the numerical simulation (black).
Since the wave packet is narrow in $\vk$-space the match is very good.
The mass gaps are chosen such that $m = 0.3 M$, while the pulse duration is set arbitrarily to $\Delta t = 1.4 \hbar/M$. 
The symmetric Dirac cone case is shown for comparison, corresponding in this setup to $m = 0,\, M=\mathcal{M}$.
The most pronounced difference is the decreasing echo signal as $\kappa\rightarrow 0$ for the hyperbolic Hamiltonian
\eqref{eq:hyperbol-Hamil} -- recall that $|\varphi_\vk^\pm\rangle$ become $\sigma_z$ eigenstates at $\kappa=0$
so that $H_1$ does not induce any transition.
Furthermore, the hyperbolic setup echo is always smaller than the Dirac cone one.
This is due to the smaller relative size of the pulse energy compared to the initial one, respectively $M$ and 
(approximately) $E^+_{\vk_0}$ from Eq.~\eqref{eq:hyperenergies},
and is reflected in the term $\frac{M}{\Mcal}$ in Eq.~\eqref{eq:transamp-hyper}.

\subsubsection{$\sigma_x$- and $\sigma_y$-pulse}
Since a mass gap pulse $\sigma_z$ loses efficiency for $\kappa_m \to 0$, let us consider a different form
of the pulse Hamiltonian $H_1$:
\be
\label{eq:hyperbol-Hamil-xy}
H = \hbar v_F \vk\cdot \vsigma + m \sigma_z + f(t)\, \Lambda  \sigma_x.
\ee
For sake of definiteness we chose $H_1 \propto \sigma_x$.  
Instead of opening a gap, this pulse shifts the band, $k_x \rightarrow k_x + f(t)\Lambda/\hbar v_F$,
while an orthogonal pulse $H_1\propto\sigma_y$ would analogously shift $k_y$.
More generally, any linear combination $H_1 \propto \alpha \sigma_x + \beta \sigma_y, \alpha,\beta\in\mathbb{R}$
could be chosen.  From Eq.~\eqref{eq:hyperbol-Hamil-xy} one has 
\ber
\hbar \Omega_\vk &=& \Lambda \sqrt{\kappa^2 + \frac{m^2}{\Lambda^2} + 2 \kappa_x + 1}, 
\label{eq:hyperbol_Omega} 
\\
h_1^\perp &=& \Lambda \langle \varphi_\vk^{\mp} \mid \sigma_x \mid \varphi_\vk^\pm \rangle 
= \frac{\Lambda}{\kappa} \left( \frac{\kappa_x}{\sqrt{1+\kappa_m^2}} \mp i \kappa_y \right),
\label{eq:hyper_hperp} 
\eer
where $\kappa = \hbar v_F k /\Lambda > 0$, $\kappa_i =\hbar v_F k_i /\Lambda \in \mathbb{R}, i=x, y $, 
and $\kappa_m = \hbar v_F k/m > 0$ as before.  The transition amplitude then reads
\be
A_\vk = -i\dfrac{ \frac{\kappa_x}{\sqrt{1+\kappa_m^2}} \mp i \kappa_y }{ \kappa\sqrt{\kappa^2 + \frac{m^2}{\Lambda^2} + 2 \kappa_x +1}} 
\sin\left( \Omega_\vk \Delta t \right),
\label{eq:transamp-hyperbol_sigx}
\ee
and analogous formulas for a $\sigma_y$-pulse are obtained
by exchanging $x\leftrightarrow y$ in Eqs.~(\ref{eq:hyperbol_Omega}-\ref{eq:transamp-hyperbol_sigx}).
In contrast to previous cases, the transition amplitude is now direction dependent.  
This reflects the preferred direction defined by the $\sigma_x$-pulse choice, 
$k_x \rightarrow k_x + f(t)\Lambda/\hbar v_F$, and implies that the wave packet propagation direction becomes important. 
Equivalently, one can fix a certain propagation direction for the wave packet and consider pulses
pointing in different directions -- in particular, the two orthogonal ones $x$ and $y$. 
This is done in Fig.~\ref{fig:hyperbolic-data}(b), which shows the $\kappa$-dependence of the echo strength
for a $\sigma_x$- and $\sigma_y$-pulse acting on a Gaussian wave packet moving along $ \hat\vk_0 =(-\hat\vk_x + \hat\vk_y)/\sqrt{2}$. 
Blue dots and green squares show simulation results respectively for a $\sigma_x$- and $\sigma_y$-pulse.
The simulation parameters are $\Delta t = 1.4 \hbar/\Lambda$, $m = 0.3 \Lambda$, and a wave packet width of $\Delta k = k_0/8$.
The light blue and light green solid lines are the corresponding analytics from Eq.~\eqref{eq:transamp-hyperbol_sigx}
-- in the latter case with $x \leftrightarrow y$.
Notice that the anisotropy of the present setup would diminish the echo of radially propagating wave packets,
such as the $\hbar$-shaped packet from Ref.~\cite{reck2017}. 
However for sufficiently strong pulses, \ie such that $\kappa\ll 1$, the anisotropic $\vk$ dependence 
of the transition amplitude plays a negligible role -- the $\sigma_x$ and $\sigma_y$ curves converge for
$\kappa\rightarrow 0$.

%%%%%%%%%%%%%%%% Parabolic bands %%%%%%%%%%%%%%%%%%%%%%%%%%%%%%%%%%%%%%%%%%%%%%%%%%%%%%%%%%%%%%%%%%%%%%%%%%%%

\subsection{Parabolic bands}
\label{subsec:parabol-bands}
% 
%%%%%%%%%%%%%%%%%%%%%%%%% FIGURE %%%%%%%%%%%%%%%%%%%%%%%%%%%%%%%%%%%%%%%%%%%%%%%%%%%%%%%%%%%%%%%%%%5
\begin{figure*}
\centering 
% \centering 
%     \def\svgwidth{\textwidth}
%     \input{./fig7.pdf_tex}  
     \includegraphics[width=\textwidth]{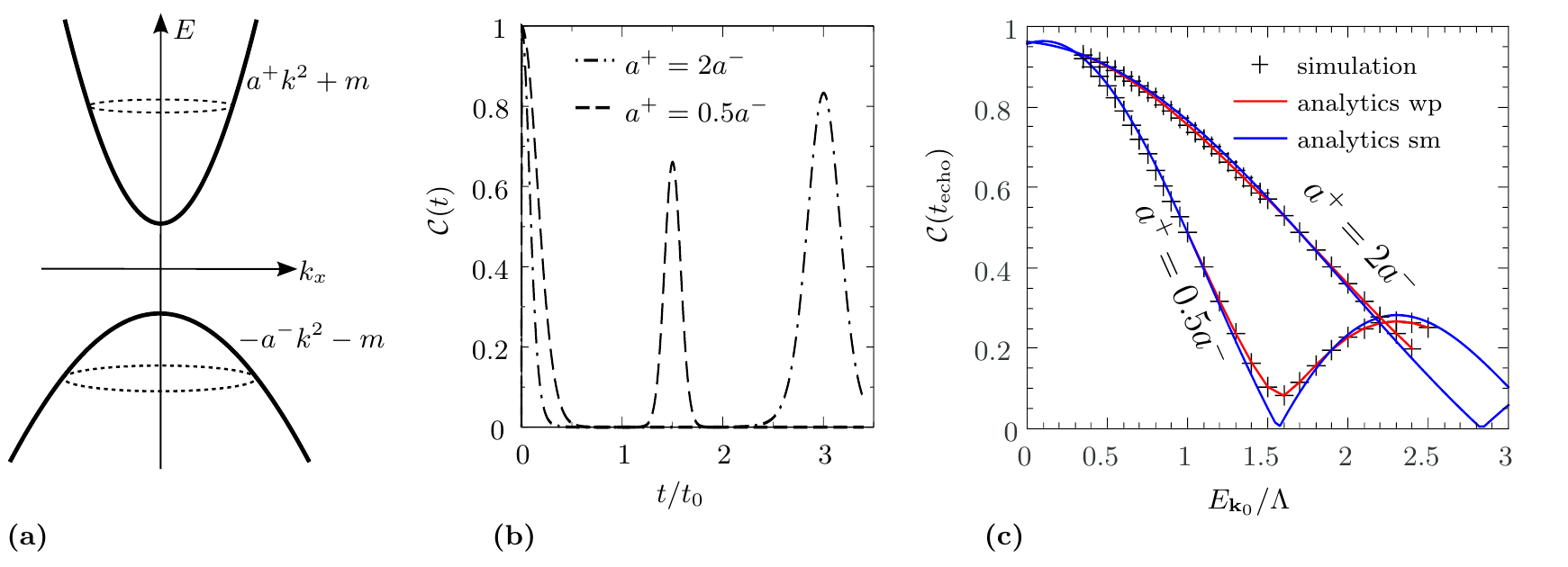} 
\centering 
\caption{
QTM for parabolic bands.
(a) Band structure (solid) of the Hamiltonian in Eq.~\eqref{eq:parabol-Hamil0} and the individual terms (dashed) of Eq.~\eqref{eq:parabol-Hamil}.
(b) Correlation $\mathcal{C}(t)$, showing that the echo time changes for going from ''fast`` to ''slow`` band compared to vice versa, according to Eq.~\eqref{eq:techo-parab}.
(c) Echo strength $\mathcal{C}(t_\echo)$ as a function of the average energy of the wave packet, obtained by simulation (black crosses), as well as using the transition amplitude of Eq.~\eqref{eq:transamp-parab} for the single mode $\vk_0$ (blue) and for the whole wave packet (red).
The data set with (mostly) smaller values corresponds to $a^+ = 0.5a^-$ and the data set with the higher echoes to $a^+ = 2a^-$.}  
\label{fig:parabolic}
\end{figure*}
%%%%%%%%%%%%%%%%%%%%%%%%% FIGURE %%%%%%%%%%%%%%%%%%%%%%%%%%%%%%%%%%%%%%%%%%%%%%%%%%%%%%%%%%%%%%%%%%5
% 
% 
To conclude our selection of representative cases, let us consider a scenario which combines chiral symmetry breaking,
as in Sec.~\ref{subsec:AsymLin}, and a $\vk$-dependent velocity within each band, as in Sec.~\ref{subsec:hyp-bands}.
We take a minimal model of two offset parabolic bands with different curvatures.
In matrix form we write
\be
H_0 = \begin{pmatrix}
	  a^+ k^2 + m & 0 \\ 0  &  - a^- k^2 - m
      \end{pmatrix}, 
\label{eq:parabol-Hamil0}
\ee
with $a^\pm>0$.  The upper and lower diagonal entries describe respectively the upper and lower bands, see Fig.~\ref{fig:parabolic}(a),
labeled by an abstract pseudospin index.
Expanding in Pauli matrices one has
\ber
H_0 
&=& 
\frac{a^+-a^-}{2} k^2  \mathbbm{1} + \left(\frac{a^++a^-}{2} k^2 + m\right) \sigma_z 
\nn \\
&=& 
\eta k^2 \mathbbm{1} + \left(\zeta k^2 + m\right) \sigma_z,
\label{eq:parabol-Hamil}
\eer
with $\eta = \frac{a^+-a^-}{2}$ and $\zeta =\frac{a^++a^-}{2} $. 
The velocities in the two bands are given by
\begin{align}
 \mathbf{v}_+(\vk) &= 2 a^+ k \\
 \mathbf{v}_-(\vk) &= -2 a^- k.
\end{align}
The QTM velocity requirement is met with $\xi_v = a^+/a^-$. 
As pulse Hamiltonian we take
\be
H_1 = f(t) \Lambda \sigma_x,
\ee
which fulfills condition \reqB~in Tab.~\ref{table_reqs}.
Furthermore, the quadratic $k$-dependence of $h_0^0$,
\be
 h_0^0 = \eta k^2, \quad
 h_1^0 = 0,
\ee
yields $\xi_h = (a^+-a^-)/a^+$,
and the echo time slightly adjusts according to Eq.~\eqref{eq:gen_t-echo}
\be
t_\echo = \left(1+ \frac{a^+}{a^-} \right) t_0 + \left(1+ \frac{ a^+-a^-}{a^-} \right) \Delta t.
\label{eq:techo-parab}
\ee
Figure \ref{fig:parabolic}(b) shows the echo times, corresponding to peaks of the correlator $\mathcal{C}(t)$, for QTMs
in which the initial and final wave packets reside respectively on the fast and slow branches of the spectrum 
(dashed-dotted line), or vice-versa (dashed line). 
The initial (mean) energy relative to the gap is $E_0/\Lambda = 0.75$ and $\Delta t\Lambda/\hbar = 1.4$.
From
\ber
h_1^\perp &=& \Lambda, 
\\
\hbar \Omega_\vk &=& \sqrt{\Lambda^2+ \left(\zeta k^2 + M_0\right) ^2 },
\eer
one has
\ber
 A_\vk &=& -i  \dfrac{\me^{-\frac{i}{\hbar} \eta k^2 \Delta t}}{\sqrt{1 + \left( \zeta k^2 + M_0\right) ^2/\Lambda^2}} 
\nn \\
&& \;\times \,\sin\left(\frac{\Lambda \Delta t}{\hbar } \sqrt{1 + \left( \zeta k^2 + M_0\right) ^2/\Lambda^2}\right).
\label{eq:transamp-parab}
\eer
Figure \ref{fig:parabolic}(c) shows the echo strength as a function of the initial (mean) energy $E_0$ of a Gaussian wave packet 
initially in the upper band for two different cases of band curvatures: $a^+ = 2 a^-$ and $a^+ = 0.5 a^-$ (while keeping $\zeta$ fixed). 
In the former case, the wave packet goes from the faster to the slower band and in the latter from the slower to the faster band. 
The pulse length is in both cases $\Delta t = 1.3 \hbar/\Lambda$. 
The curve in panel (c) that decays faster corresponds to the transition from slower to faster bands. 
The reason of the smaller echoes for the same energies $E_{\vk_0}$ in this case ($a^+ = 0.5 a^-$) is that the initial $k_0$ (and therefore $\Omega_{\vk_0}$) is larger due to the smaller band curvature than in the other case ($a^+ = 2 a^-$).
Thus, transition amplitude of Eq.~\eqref{eq:transamp-parab} is smaller and the overall echo is diminished.
This could be already seen in Fig.~\ref{fig:parabolic}(b), where not only the echo times but also the echo strength are quite different in the two cases.

We conclude that good echoes could be achieved in systems
whose Hamiltonians share the qualitative features of Eq.~\eqref{eq:parabol-Hamil}, \eg direct gap semiconductors 
or bilayer graphene.

%%%%%%%%%%%%%%%%%%%%%%%%%%%%% EFFECT OF PERTURBATIONS On THE QTM %%%%%%%%%%%%%%%%%%%%%%%%%%%%%%%%%%%%%%%%%%%%%%%%%%%

\section{Perturbations}
\label{sec:Perturbations}

To deal with perturbations, in particular the effect of an additional electric or magnetic field, we consider the QTM protocol from a slightly different point of view,
and rely on concepts belonging to Loschmidt echo theory \cite{calvo2008,goussev2012b}.

\subsection{Echo fidelity and general approach}
\label{subsec:pert:gen}
A propagation backward in time is mathematically equivalent to inverting the sign of the Hamiltonian, 
$H\to-H$, in the time-evolution operator.
For the Hamiltonians considered thus far, this is effectively achieved by the QTM, which is nothing other 
than a population reversal protocol.
In the real world there are however numerous limitations to the practical implementation of such a program.
Even in the absence of interactions -- whose role will be discussed later in Sec.~\ref{sec:manybody} --
it is difficult to effectively invert the time evolution generated by 
the full Hamiltonian $H$, since not all of its parameters are usually under complete control or can be tuned in a time-resolved manner.
In practice, the QTM pulse will effectively invert only the time evolution due to a certain part of $H$,
the one which can be characterized and controlled precisely.  
Let us start by taking the latter of the general form \eqref{eq:H0-gen} 
\begin{equation}
H_0 = h_0^0(\vk) \mathbbm{1} + \bh_0(\vk)\cdot\vsigma, 
\end{equation}
and considering the corresponding time evolution from $t=0$ to some time $t=t_0 + \Delta t + t^\prime$ after the pulse.
For the following discussion it is convenient to write the pseudospin-dependent part as
\be
\bh_0(\vk)\cdot\vsigma = h_0(\vk)\,\hbn_\vk \cdot \vsigma
\ee
with $\hbn_\vk$ a $\vk$-dependent unit vector.
The $H_0$ eigenstates are denoted $|\varphi_\vk^s\rangle,\,s=\pm$,
and the operator (of unit modulus) mapping $|\varphi_\vk^s\rangle$ to $|\varphi_\vk^{-s}\rangle$ is given by 
\begin{equation}
 \hbm_\vk\cdot\vsigma  |\varphi_\vk^s\rangle = |\varphi_\vk^{-s}\rangle, 
\label{eq:transop-pert} 
\end{equation}
with the unit vector $\hbm_\vk \perp \hbn_\vk$. 
Geometrically, $\hbm_\vk\cdot\vsigma$ defines a $\pi$ rotation on the pseudospin Bloch sphere around the axis $\hbm_\vk$ 
\begin{equation}
\hbm_\vk\cdot\vsigma = \cos \pi/2 \mathbbm{1} + \hbm_\vk\cdot\vsigma  \sin \pi/2 = \me^{i \frac{\pi}{2}  \hbm_\vk\cdot\vsigma}.
\end{equation}
If $\hbn_\vk$ lies in an arbitrary plane for all values of $\vk$, as is the case for the Dirac Hamiltonians
in Secs.~\ref{subsec:pristine} and \ref{subsec:AsymLin},
then there is a unique direction $\hbm \perp \hbn_\vk \, \forall \vk$, 
and the transition operator becomes $\vk$-independent
\footnote{Otherwise $\hbm_\vk$ is not unique but lies in the plane perpendicular to $\hbn_\vk$.}
% See also the discussion in the caption of Fig.~\ref{fig:???}}.
In the following, we take as ideal pulse Hamiltonian
\begin{equation}
H_\pulse = M f(t) \hbm\cdot\vsigma,\quad M\in\mathbb{R}. 
\label{eq:pulseHamil-pert}
\end{equation}
Given the time-evolution of an initial eigenstate, we focus only on the ``backward'' propagating part generating the echo. 
This is done by projecting the time-evolution of $|\varphi_\vk^s\rangle$ onto its chiral counterpart $|\varphi_\vk^{-s}\rangle$:
\begin{align}
\mathcal{F}_\vk(t)&=|\langle\varphi_\vk^{-s} \mid U(t,0) \mid \varphi_\vk^s\rangle|^2 \nonumber \\
&= |\langle\varphi_\vk^{-s} \mid \me^{-\frac{\iu}{\hbar} H_0 t^\prime } \me^{-\frac{\iu}{\hbar} (H_0+H_\pulse)\Delta t } \me^{-\frac{\iu}{\hbar} H_0 t_0 } \mid \varphi_\vk^s\rangle|^2,
\label{eq:echofid0-noPert}
\end{align}
for times after the pulse, with $t=t_0+\Delta t + t^\prime$.
We dub the quantity $\mathcal{F}_\vk(t)$ the ``echo fidelity'' (for a single $\vk$-mode), in analogy with the standard fidelity measure.
From the expansion in  Eq.~\eqref{eq:timeevo1} follows
\be
\mathcal{F}_\vk(t) = |\langle\varphi_\vk^{-s} \mid \me^{-\frac{i}{\hbar} H_0 t^\prime }  \hbm\cdot\vsigma A_\vk \me^{-\frac{i}{\hbar} H_0 t_0 } \mid \varphi_\vk^s\rangle|^2,
\ee
where all terms that do not lead to a transition between bands are neglected.  The transition amplitude $A_\vk$ is defined in Eq.~\eqref{eq:gen-transamp}.
Rewriting $\langle\varphi_\vk^{-s}|$ via Eq.~\eqref{eq:transop-pert} and after some manipulations one finds
\begin{align}
\mathcal{F}_\vk (t)
&=
|  A_\vk|^2\,\langle\varphi_\vk^{s} \mid \hbm\cdot\vsigma\,\me^{-\frac{i}{\hbar} H_0 t^\prime } \,\hbm\cdot\vsigma\,
\me^{-\frac{i}{\hbar} H_0 t_0 } \mid \varphi_\vk^s\rangle |^2
\nn \\
&= 
|A_\vk|^2\,\langle\varphi_\vk^{s} \mid\me^{-\frac{i}{\hbar}  \hbm\cdot\vsigma\, H_0 \,\hbm\cdot\vsigma t^\prime }   
\me^{-\frac{i}{\hbar} H_0 t_0 } \mid \varphi_\vk^s\rangle|^2 
\nn \\
&= 
| A_\vk|^2\,\langle\varphi_\vk^{s} \mid\me^{-\frac{i}{\hbar}  (h_0^0 \mathbbm{1}-\mathbf{h}_0\cdot\vsigma ) t^\prime }  
\me^{-\frac{i}{\hbar} (h_0^0 \mathbbm{1}+\mathbf{h}_0\cdot\vsigma ) t_0 } \mid \varphi_\vk^s\rangle|^2.
\label{eq:echofid-nopert}
\end{align}
The sign of $\bh_0$, the pseudospin-dependent part of the Hamiltonian, is effectively inverted which is equivalent to a propagation backwards in time. 
The pseudospin-independent term $h_0^0\mathbbm{1}$ on the other hand does not change sign, but as long as it is in agreement with the requirements for the QTM of Subsec.~\ref{subsec:gentransamp}, it will only change the echo time $t_\echo$ according to Eq.~\eqref{eq:gen_t-echo}.  See also App.~\ref{app:t_echo}, 
where $t_\echo$ is derived in a similar way.
The term $\propto \mathbbm{1}$ will however cause problems 
in the presence of a position-dependent potential, as discussed below.

We now add to $H_0$ a position- and pseudospin-dependent potential $\mathcal{V}(\br)$,
\begin{align}
\mathcal{V}(\br) 
&=
V_0(\br) \mathbbm{1} + \mathbf{V}(\br) \cdot \vsigma 
\nn \\
&= 
V_0(\br) \mathbbm{1} + V_l(\br) \hbl_{\vq}
\cdot \vsigma.
\end{align}
Throughout the rest of the section $\vk, \br$ will in general be the non-commuting momentum and position operators,
and we will specify the relevant matrix elements on a case-by-case basis.  
The eigenenergies and eigenstates of the full static Hamiltonian $H_0+\mathcal{V}$ 
will be denoted by $\epsilon_n^s$ and $|\phi_n^s\rangle, \,s=\pm$, and the transition operator is defined as in Eq.~\eqref{eq:transop-pert} by 
\begin{equation}
 \hbm_n  \cdot \vsigma | \phi_n^s\rangle =  |\phi_n^{-s}\rangle.
\end{equation}
The pulse Hamiltonian is again homogeneous and $n$-independent, as in Eq.~\eqref{eq:pulseHamil-pert}.

%%%%%%%%%%%%%%%%%%%%%%%%%%%%%%%%%%%%%%%%%%%%%%%%%%%%%%%%%%%%% COSIMO V13: ab hier %%%%%%%%%%%%%%%%%%%%%%%%%%%%%%%%%%%%%%%%%%%%%%%%%%%%%%%%%%%

We proceed as before and compute the echo fidelity for the full Hamiltonian
\be
H = H_0 + H_\pulse + \mathcal{V}(\br).
\ee
The echo fidelity of Eq.~\eqref{eq:echofid0-noPert} now reads
\ber
\mathcal{F}_n(t)
&=&
|\langle \phi_n^{-s} \mid U(t_\echo,0) \mid \phi_n^s\rangle|^2 
\nn \\
&=& 
|\langle\phi_n^{-s} \mid \me^{-\frac{i}{\hbar} (H_0+\mathcal{V}) t^\prime } \me^{-\frac{i}{\hbar} (H_0+\mathcal{V}+H_\pulse)\Delta t }
\nn \\ 
&& \quad\quad\quad\quad\quad
\times \me^{-\frac{\iu}{\hbar} (H_0+\mathcal{V}) t_0 } \mid \phi_n^s\rangle|^2.
\label{eq:echofid0-pert}
\eer
The simple expansion \eqref{eq:timeevo1} for the time evolution operator during the pulse is no longer valid, 
the Hamiltonian being a function of both momentum and position operators.  
Nevertheless we can still expand it in terms of Pauli matrices
\be
\label{eq:pulse_pauli}
 \me^{-\frac{i}{\hbar} (H_0+\mathcal{V}+H_\pulse)\Delta t } = \gamma_0(\vq,\vk; \Delta t) \mathbbm{1} 
+ \boldsymbol{\gamma}(\vq,\vk; \Delta t)\cdot \vsigma,
\ee
with $\gamma_0, \boldsymbol{\gamma}$ functions of the position and momentum operator and of the pulse duration $\Delta t$.  
The term $\boldsymbol{\gamma}\cdot\vsigma$
reshuffles both orbital ($n$) and helicity ($s$) quantum numbers,
\ber
\boldsymbol{\gamma}\cdot\vsigma 
&=& 
\sum_{n'n, s's}\, \gamma_{n'n}^{s's}|\phi_{n'}^{s'}\rangle\langle\phi_n^s|
\nonumber\\
\label{eq:reshuffle}
&=& \gamma_{nn}^{-ss}|\phi_{n}^{-s}\rangle\langle\phi_n^s| + 
{\sum_{n'n, s's}}^*\, \gamma_{n'n}^{s's}|\phi_{n'}^{s'}\rangle\langle\phi_n^s|.
\eer
Here, the summation $\sum^*_{n'n, s's}$ runs over all states but the one singled out,
which connects $|\phi_n^s\rangle \leftrightarrow |\phi_n^{-s}\rangle$ and is thus the only one contributing to the echo. 
\footnote{Note that here, we assume that only the velocity from $|\phi_n^{-s}\rangle$ can be such 
that it inverts the motion of $|\phi_n^{-s}\rangle$ before the pulse. 
In special systems, also $|\phi_{n^\prime}^{s^\prime}\rangle$ with $n^\prime\neq n$ and $s^\prime = \pm$ might be able to do the same. 
However, we do not consider such an extreme case here.}
This term is in general small compared to the rest of the sum. 
It can however be the dominant one \eg for short pulses 
not leading to considerable orbital changes.

The echo fidelity is computed as usual neglecting the term $\propto \mathbbm{1}$ in Eq.~\eqref{eq:pulse_pauli}, 
as it does not lead to transitions $|\phi_n^s\rangle \leftrightarrow |\phi_n^{-s}\rangle$, 
and plugging the expression \eqref{eq:reshuffle} into Eq.~\eqref{eq:echofid0-pert},
\begin{widetext}
\ber
\mathcal{F}_n(t)
&=&
|\langle\phi_n^{-s}\mid\me^{-\frac{i}{\hbar}\left(H_0+\mathcal{V}\right)t'}\mid\phi_n^{-s}\rangle
\,\gamma_{nn}^{-ss}(\Delta t)\,\langle\phi_n^s\mid\me^{-\frac{i}{\hbar}\left(H_0+\mathcal{V}\right)t_0}\mid\phi_n^s\rangle|^2
\nonumber\\
&=&
|\gamma_{nn}^{-ss}(\Delta t)|^2\,\,
|\langle\phi_n^s\mid
\me^{-\frac{i}{\hbar}
\hbm\cdot\vsigma
\left(H_0+\mathcal{V}\right)\hbm\cdot\vsigma 
t^\prime}
\,
\me^{-\frac{i}{\hbar}\left(H_0+\mathcal{V}\right)t_0}\mid\phi_n^s\rangle|^2.
\label{eq:echofid-pert}
\eer
\end{widetext}
In the second line we performed manipulations analogous to those leading to Eq.~\eqref{eq:echofid-nopert}.
%%%%%%%%%%%%%%%%%%%%%%%%% FIGURE %%%%%%%%%%%%%%%%%%%%%%%%%%%%%%%%%%%%%%%%%%%%%%%%%%%%%%%%%%%%%%%%%%5
\begin{figure*}
 \centering 
%  \centering 
%     \def\svgwidth{0.8\textwidth}
%     \input{./fig8.pdf_tex}
     \includegraphics[width=\textwidth]{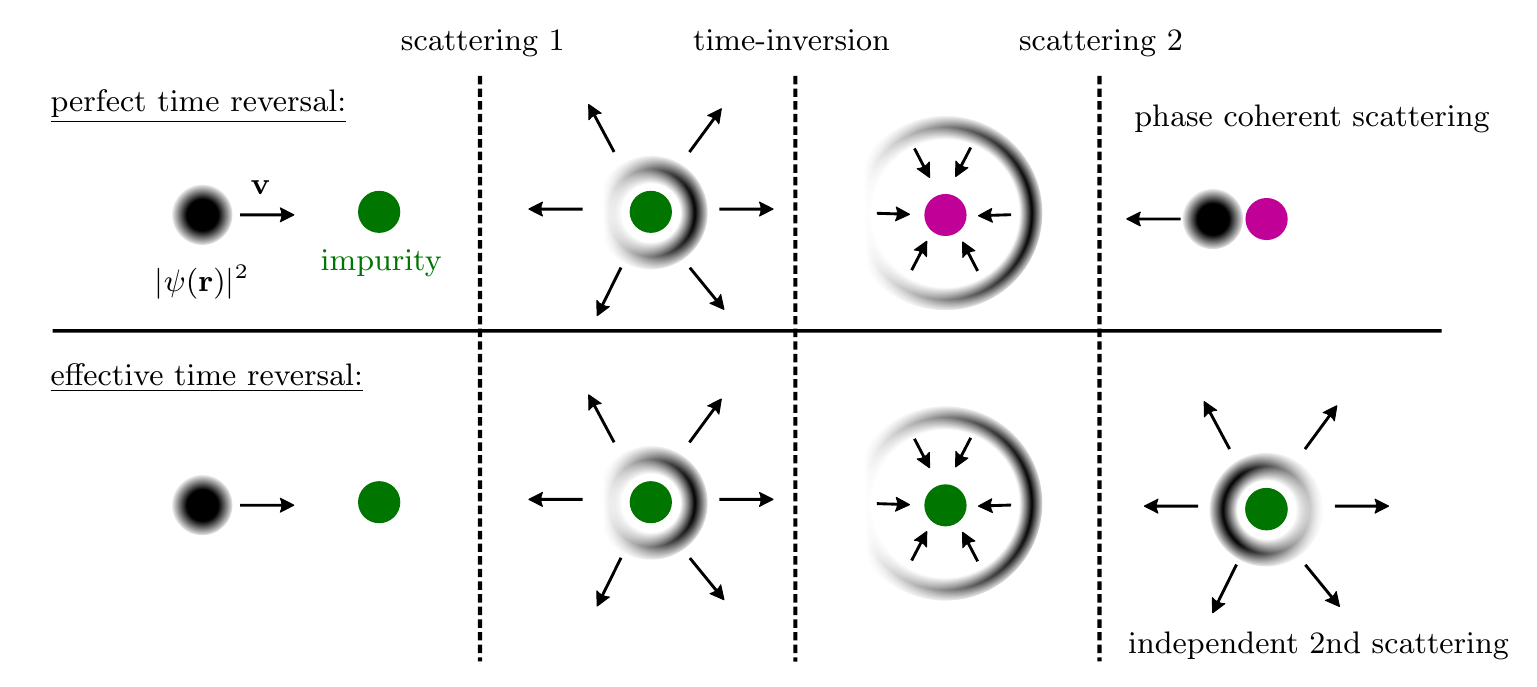} 
\caption{
Difference between a perfect (hypothetical) time reversal (upper panels) and our effective one (lower panels) by the QTM in the presence of disorder. A perfect time reversal changes the sign of every term in the Hamiltonian, indicated by the change of color of the impurity from green to magenta, which leads to a phase coherent backscattering. On the other hand, the QTM pulse does not effectively switch the sign of the impurity potential in the echo fidelity due to Eq.~\eqref{eq:disorder}, which is why the second scattering process after the inversion of propagation direction is independent from the first one, which is why it does not lead to the initial shape of the wave function.
} \label{fig:disorder} 
 \centering 
\end{figure*}
%%%%%%%%%%%%%%%%%%%%%%%%% FIGURE %%%%%%%%%%%%%%%%%%%%%%%%%%%%%%%%%%%%%%%%%%%%%%%%%%%%%%%%%%%%%%%%%%5
The important message is that
a change of sign in the time-evolution operator after the pulse, and thus an effective time-inversion, is only achieved if
\begin{equation}
[\hbm\cdot \vsigma] \;(H_0+\mathcal{V})\; [\hbm\cdot \vsigma] = -  (H_0+\mathcal{V}).
\label{eq:Req-pert}
\end{equation}
Most importantly, this means that terms in $H_0+\mathcal{V}$ proportional to $\mathbbm{1}$ are in general not effectively time-inverted by our mechanism.
We emphasize that this conclusion has a geometrical origin and is thus general,
independent of the specific form of the Hamiltonian.

As a known example \cite{reck2017} where the potential destroys the time-reversal, take a generic Dirac system with a mass pulse ($\propto \sigma_z$) as in Subsec.~\ref{subsec:pristine}, 
in the presence of pseudospin independent disorder 
\be
  \mathcal{V}^\imp(\br) = V^\imp_0(\br) \mathbbm{1}.
\ee
Disorder of this type is detrimental to the echo,
since the $\mathcal{V}^\imp(\br)$-induced dynamics cannot be effectively time-inverted by the QTM pulse: 
\be
\sigma_z  \mathcal{V}^\imp(\br) \sigma_z =  \mathcal{V}^\imp(\br)\neq -\mathcal{V}^\imp(\br).
\label{eq:disorder}
\ee
Even if the dynamics of an injected wave packet stays phase-coherent,
Eq.~\eqref{eq:disorder} means that the randomness which the potential landscape transmits to the phases
of each $\vk$-mode on the forward time-evolution cannot be removed by the QTM on the ``backward'' branch, see also Fig.~\ref{fig:disorder}.
For an extended discussion of such a dephasing mechanism and its consequences for the echo
see Ref.~\cite{reck2017}, where the scenario is treated in detail.

We now consider the more general case of band structures of $H_0$ which are not chiral symmetric but fulfill the weaker requirement of Eq.~\eqref{eq:velrequirement} demanded in the clean system in Sec.~\ref{sec:genmod}.
With Eq.~\eqref{eq:Req-pert} we see that in those systems any potential $\mathcal{V}(\vq)\neq const$ is detrimental to the QTM principle. 
The reason is that in these systems $H_0$ needs to have a term $h_0^0\mathbbm{1}$, because otherwise, the band structure would be chiral symmetric. 
On the other hand, we have seen in Eqs.~\eqref{eq:echofid-pert} and \eqref{eq:echofid-nopert} that these terms ($\propto \mathbbm{1}$) are not time-inverted by the QTM pulse. 
In those systems, an echo was still possible without $\mathcal{V}(\vq)$, not at time $t^\prime=t_0$ but at some other time, because all $\vk$-modes still lose their energy-dependent phase at exactly the same time. 
With $\mathcal{V}(\vq)$ on the other hand, the position operators will change the momentum of the wave packet, $\vk$ not being a good quantum number anymore.
Although this change of $\vk$ might be time-inverted by the QTM pulse, the initial momentum is recovered only exactly at $t^\prime = t_0$, but the phases due to $H_0(\vk)$ will not cancel at that point in time -- they did not even cancel in the clean system. 
Therefore, no true echo of the initial wave packet is to be expected. 

Note that in principle there could be systems
where $H_0(\vk)$ and $\mathcal{V}(\vq)$ are ``matched'', in the sense that although the full system is not effectively time-inverted, 
all modes $|\phi_n^s\rangle$ cancel their energy-dependent phases at an $n$-independent time $t^\prime\neq t_0$.
This would be the generalization of the clean case considered above in Secs.~\ref{sec:genmod} and \ref{sec:twoband-examples}, 
where the band structure is not chiral symmetric and 
one branch is stretched with a $\vk$-independent constant as compared to the other one, compare Eq.~\eqref{eq:velrequirement}.
This appears however as a rather extreme case, and it is therefore not further discussed.

To achieve high fidelity echoes also for wave packets consisting of many eigenstates $|\phi_n^s\rangle$, 
apart from the fact that there should not be a $\mathbbm{1}$-term in $H_0$ nor in $\mathcal{V}$, 
one has to additionally make sure that the pulse $\sim \hbm\cdot \vsigma$ is essentially the transition operator $\hbm_n\cdot \vsigma$ for all relevant $n$ of the wave packet.
In the clean system, this would be requirement \reqB ~of Subsec.~\ref{subsec:gentransamp}, which was the reason why the echo in a hyperbolic band structure is always weaker than in graphene with comparable parameters (see Fig.~\ref{fig:hyperbolic-data}(a)).
For a general system, this requirement concerning $\hbm$ and $\hbm_n$ might not be possible, since this would necessarily mean that $\hbn_\vk$ and $\hbl_\vq$, \ie the pseudospin-directions of $H_0(\vk)$ and $\mathcal{V}(\vq)$, respectively, are in the same plane for all $\vk,\vq$.
However, if this is the case (see graphene in a magnetic field in Subsec.~\ref{subsec:magnetic} as an example), we can even make quantitative predictions of the echo strength.
Take the static part of the Hamiltonian to be 
\be
 H_0(\vk) = h_0(\vk) \hbn_\vk\cdot \vsigma,\quad
 \mathcal{V}(\vq) =  V_l(\br) \hbl_\vq \cdot \vsigma
\ee
and $H_\pulse$ as in Eq.~\eqref{eq:pulseHamil-pert},
with $\hbn_\vk, \hbl_\br \perp \hbm$ for all $\vk, \vq$ \footnote{As before, $\hbn_\vk$ and $\hbl_\vq$ 
need can be arbitrarily pointed with respect to each other.}.
As shown in App.~\ref{app:expansion-time-evo-pulse}, we then can exactly expand the time evolution operator during the pulse acting on an eigenstate of $H_0+\mathcal{V}$:
\begin{align}
 \me^{-\frac{\iu}{\hbar} (H_0+\mathcal{V}+H_\pulse)\Delta t } \;| \phi_n^\pm\rangle  =  B^\pm_n  \;| \phi_n^\pm\rangle  + A_n  \;| \phi_n^\mp\rangle 
\label{eq:echoFid-V-1}
\end{align}
where the transition amplitude reads
\begin{equation}
 A_n = \frac{-\iu}{\sqrt{1+\epsilon_n^2/M^2}}\;\sin\left(\frac{\Delta t M }{\hbar}\sqrt{1+\epsilon_n^2/M^2 }\right).
\label{eq:transamp-pert}
\end{equation} 
We thus get a generalized form of Eq.~\eqref{eq:echofid-nopert},
\begin{widetext}
\be
\mathcal{F}_n(t)
=
|A_n|^2\, 
|\langle\phi_n^s \mid\me^{+\frac{i}{\hbar}  
(H_0+\mathcal{V})
t^\prime }
\,
\me^{-\frac{\iu}{\hbar} (H_0+\mathcal{V}  ) t_0 } \mid \phi_n^s\rangle|^2.
\label{eq:echofid-pert-perpendicular}
\ee
\end{widetext}
This is very similar to the result without $\mathcal{V}$ from Eq.~\eqref{eq:echofid-nopert}, with the obvious difference that now the fraction $|A_n|^2$ of any $n$-mode is effectively time-inverted instead of $|A_\vk|^2$ of any $\vk$-mode.
Thus, no matter the strength of $\mathcal{V}$, effective time-reversal can be achieved.

We conclude this general section about perturbations $\mathcal{V}(\vq)$ by discussing the case when $H_0$ is of the special form that there exists an $\hbm$ such that $[\hbm\cdot \vsigma] \;H_0\; [\hbm\cdot \vsigma] = -H_0$, $\forall \vk$, as for example in Dirac systems like graphene.
The main result, which is the requirement of Eq.~\eqref{eq:Req-pert}, then simplifies and can be separated into three cases:
\begin{enumerate}
 \item[\casepertA] $[\hbm\cdot \vsigma] \;\mathcal{V}(\vq)\; [\hbm\cdot \vsigma] = \mathcal{V}(\vq)$,
 \item[\casepertB] $[\hbm\cdot \vsigma] \; \mathcal{V}(\vq) \; [\hbm\cdot \vsigma] \neq \pm  \mathcal{V}(\vq)$,
 \item[\casepertC] $[\hbm\cdot \vsigma] \;\mathcal{V}(\vq)\; [\hbm\cdot \vsigma] = -\mathcal{V}(\vq)$.
\end{enumerate}
In case \casepertA, the echo will decline for increasing $\mathcal{V}$ as compared to the case $\mathcal{V}=0$.
Examples for this case are graphene with disorder as discussed above or graphene in a homogeneous, electric field, see Subsec.~\ref{subsec:electric}.
In case \casepertB, an effective time-reversal is in principle possible. 
The problem is that eigenstates $|\phi_n^s\rangle$ of $H_0 + \mathcal{V}$ are not necessarily efficiently mapped by $H_\pulse$ to their counterparts $|\phi_n^{-s}\rangle$ with the same orbital quantum number $n$, which is reflected 
in Eq.~\eqref{eq:echofid-pert} by the quantity $|\gamma_{nn}^{-ss}|$.
The exact effect of $\mathcal{V}$ in this case has to be evaluated for each system individually.

In the special case \casepertC, the transition amplitude $A_n$ is obtained analytically in Eq.~\eqref{eq:transamp-pert} and differs only from the case $\mathcal{V}=  0 $ in the sense that we have to consider the combined orbital quantum number $n$ instead of the momentum $k$.
Most importantly that means that high fidelity echoes can be in principle achieved no matter how strong $\mathcal{V}(\vq)$ is. 
An example of case \casepertC ~is graphene in a magnetic field as discussed in Subsec.~\ref{subsec:magnetic}.

%%%%%%%%%%%%%%%% Electric field %%%%%%%%%%%%%%%%%%%%%%%%%%%%%%%%%%%%%%%%%%%%%%%%%%%%%%%%%%%%%%%%%%%%%%%%%%%%

\subsection{Homogeneous electric field}
\label{subsec:electric}
%
%%%%%%%%%%%%%%%%%%%%%%%%% FIGURE %%%%%%%%%%%%%%%%%%%%%%%%%%%%%%%%%%%%%%%%%%%%%%%%%%%%%%%%%%%%%%%%%%5
\begin{figure}
 \centering 
% \begin{subfigure}[b]{\textwidth}
%  \centering 
%     \def\svgwidth{\columnwidth}
%     \input{./fig9.pdf_tex}
     \includegraphics[width=\columnwidth]{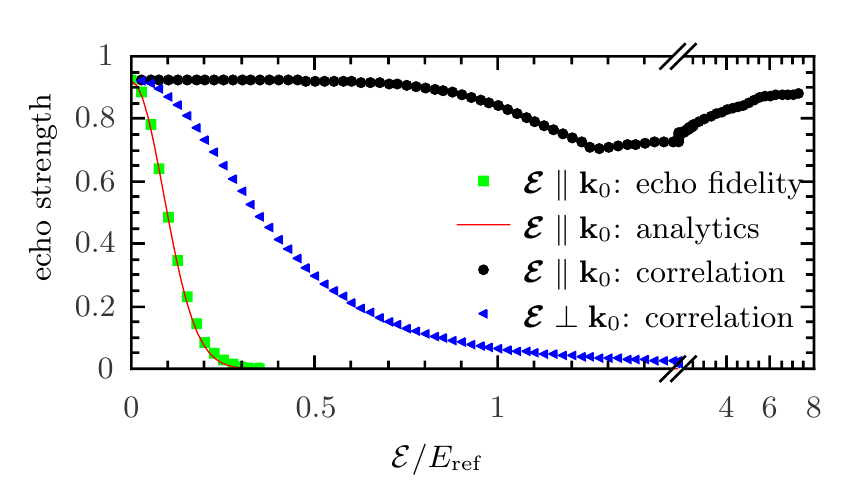} 
\caption{
Dependence of the echo strength on the electric field. 
The echo decreases in most cases monotonously, since the QTM cannot effectively time-invert the system. 
This decrease is visible in the correlation $\mathcal{C}$ in the case $\mathbf{E}_0\perp\vk_0$ (blue triangles) as well as in the echo fidelity defined in Eq.~\eqref{eq:echofid0-pert} also in the parallel case. 
The green squares indicate the numerically obtained echo fidelity for a Gaussian wave packet and the red line is the corresponding analytical calculation, assuming a constant $\vk$-shift as in Eq.~\eqref{eq:Efield-shiftKspace}. 
On the other hand, the correlation for the parallel case (black dots) is over a wide range independent of the electric field, as explained in the text.
} \label{fig:E-field-data} 
 \centering 
\end{figure}
%%%%%%%%%%%%%%%%%%%%%%%%% FIGURE %%%%%%%%%%%%%%%%%%%%%%%%%%%%%%%%%%%%%%%%%%%%%%%%%%%%%%%%%%%%%%%%%%5
%
\begin{figure*}
 \centering 
%  \centering 
%     \def\svgwidth{0.99\textwidth}
%     \input{./fig10.pdf_tex}  
     \includegraphics[width=0.99\textwidth]{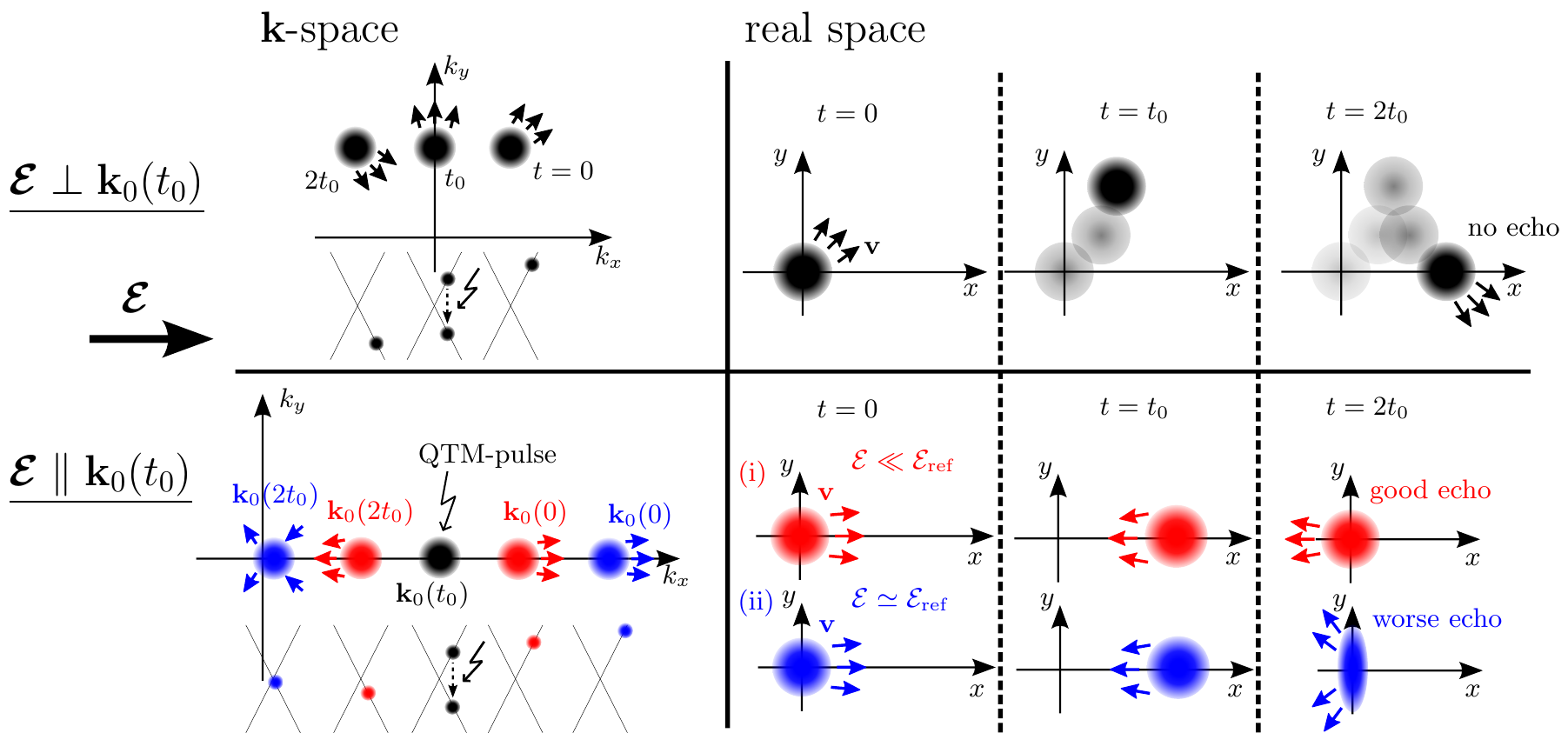} 
 \centering 
\caption{
Sketch to explain the qualitative difference between $\Ecalb\perp\vk_0(t_0)$ (upper panels) and $\Ecalb\parallel\vk_0(t_0)$ (lower panels). 
The $\vk$- and real space propagation of Gaussian wave packets in an electric field is shown, causing a linear shift in momentum space. 
In the case $\Ecalb\perp\vk_0(t_0)$, this shift leads to a change of propagation direction which is why the wave packet does not return to its initial position and therefore, a suppressed echo is expected for sufficiently large electric field.
For $\Ecalb\parallel\vk_0(t_0)$, the propagation direction of $\vk_0$ does not change which is why echoes with high fidelity are to be expected  for $\Ecal\ll\Ecal_\tref$ (red). 
Only for $\Ecal\simeq\Ecal_\tref$ (blue), spreading effects due to the $\vk$-space width $\Delta k$ diminish the echo, which become important close to the Dirac point (see also Fig.~\ref{fig:E-field-data}).
After the QTM-pulse, only the parts of the wave function which switched bands are shown, because the others do not contribute to the echo.
}  \label{fig:Efield-wavepackets}
\end{figure*}
%
%%%%%%%%%%%%%%%%%%%%%%%%% FIGURE %%%%%%%%%%%%%%%%%%%%%%%%%%%%%%%%%%%%%%%%%%%%%%%%%%%%%%%%%%%%%%%%%%5
%
%
Now, we verify the general results of Subsec.~\ref{subsec:pert:gen} for the important case of a Dirac system in a static and homogeneous electric field.
The corresponding Hamiltonian reads
\begin{equation}
 H = \hbar v_F \vk\cdot\vsigma + e \Ecalb \cdot\br \mathbbm{1} + f(t) M\sigma_z.
\end{equation}
Due to the pseudospin-independence of the electric field term, the requirement \eqref{eq:Req-pert} is not met, but instead we have
\begin{equation}
 \sigma_z e \Ecalb \cdot\br \mathbbm{1} \sigma_z =  e \Ecalb \cdot\br \mathbbm{1},
\end{equation}
\ie we are in case \casepertA ~described in the summary of Subsec.~\ref{subsec:pert:gen} and thus, the action of the electric field cannot be effectively time-inverted.
This is best seen by looking at the time-evolution of a state $\psi$ in reciprocal space due to the electric field,
\begin{equation}
   \exp\left(-\frac{\iu}{\hbar} e\Ecalb\cdot \br \, t \right) \psi(\vk) = \psi\left(\vk + \frac{e\Ecalb}{\hbar}t  \right).
\label{eq:Efield-shiftKspace}
\end{equation} 
For Dirac quasiparticles the linear shift in momentum implies no change of the magnitude of the velocity $|\mathbf{v}|$\;=\;$v_F$, 
since the velocity does not depend on $|\vk|$, but only a change of direction:
\begin{equation}
 \mathbf{v}^\pm(\vk) = \pm v_F \frac{\vk}{|\vk|}. 
\label{eq:vel-linbands}
\end{equation}
The $\vk$-shift is pseudospin-independent, \ie it is equal for both particle- and hole-like states.
Therefore it cannot be reverted by a pulse inducing particle-hole transitions.
The numerical simulations confirm the decay of the echo fidelity for larger electric fields, as shown in Fig.~\ref{fig:E-field-data},
where a Gaussian wave packet evolves in presence of an electric field pointing in $x$-direction.
The mean wave vector of the initial wave packet is chosen to depend on the strength of the electric field,
$\vk_0(t=0) = \vk_0(t=0, \Ecalb)$: The initial value is chosen such as to have equal mean wave vectors at the pulse time $t_0$
independent of the electric field strength, $\vk_0(t=t_0)\,\forall\Ecalb$ \footnote{Therefore, the transition amplitude $A(\vk_0(t_0))$ is the same in all cases, independent of the electric field}.  
This way the pulse effect is in all cases the same, and the impact of the varying electric field on the forward and backward
time-evolution is singled out.  Furthermore we have $\hbar v_F k_0(t_0)/M = 0.4$, a wave packet $\vk$-space width of 
$\Delta k = k_0(t_0)/8$ and a pulse duration $\Delta t M/\hbar = \pi/2$.  
The electric field is normalized to $\Ecal_\tref=\hbar k_0(t_0) /(et_0)$, which is the electric field needed to change the momentum in time $t_0$ until the pulse by its overall magnitude of $k_0(t_0)$. 
For instance, for wave packets with $k_0(t_0)\simeq 100\,$meV and a pulse time of $t_0\simeq 1\,$ps, 
the reference electric field is $\Ecal_\tref\simeq10^5\,$V/m.
Green squares in Fig.~\ref{fig:E-field-data} show the numerically computed echo fidelity, Eq.~\eqref{eq:echofid0-pert},
perfectly matching its analytical counterpart (red line) \footnote{The analytic curve is given by the overlap of two Gaussian wave packets in $\vk$-space, which overlap less and less for stronger electric fields.}.  As anticipated, the $\vk$-space overlap between initial
and time-mirrored packets decreases with increasing electric field strength due to Eq.~\eqref{eq:Efield-shiftKspace}.

For the correlator $\mathcal{C}$, Eq.~\eqref{eq:spaceCorr}, the picture is slightly different as sketched in Fig.~\ref{fig:Efield-wavepackets}.
For this observable only real space overlaps of the (initial and echoed) wave packets moduli are important. 
Consider first the $\vk_0\nparallel\Ecalb$ case, and in particular the extreme situation 
$\vk_0\perp\Ecalb$, blue triangles in Fig.~\ref{fig:E-field-data}.
The electric field-induced $\vk$-shift changes the velocity direction, which leads to a different trajectory of the wave packet after the pulse as compared to before (see upper panels in Fig.~\ref{fig:Efield-wavepackets}), which is why %disrupts the shape of the initial wave packet:
% The initial spread of the constituent $k_y$-modes (of width $\Delta k$) increases both before and after the QTM pulse has acted, and
 the echo strength measured by $\mathcal{C}$ declines monotonically as a function of the electric field strength.
On the other hand, if the wave packet consisted only of $\vk$-modes parallel to $\Ecalb$ the propagation direction would not change and the echo would not be affected.

The most interesting situation is however that
of a wave packet with finite width $\Delta k$, and such that $\vk_0$ and $\Ecalb$ are aligned (see black dots in Fig.~\ref{fig:E-field-data} and lower panels in Fig.~\ref{fig:Efield-wavepackets}).
The situation is qualitatively different from and more complex than either of the above extreme cases.  
First, we expect in general the correlator $\mathcal{C}$ to be practically independent of the electric field strength as long as $\Ecal\ll\Ecal_\tref$ (red parts in Fig.~\ref{fig:Efield-wavepackets}), which ensures that the condition  $\Delta k \ll k_0$ valid for the initial wave packet holds up for all times till $t_\echo$.  
In this case the $y$-velocity spread is de facto unaffected by the $\vk$-shift, so that the injected wavepacket propagates back to its initial position without changing its profile.
For stronger fields, $\Ecal\approx\Ecal_\tref$, the wave packet will come close to the Dirac point in $\vk$-space at some point before the echo (see blue parts in Fig.~\ref{fig:Efield-wavepackets}). 
Since there, its velocity distribution changes dramatically, the wave packet spreads and its real space shape is altered.
Note that this is independent of whether the electric field $\Ecalb$ is parallel or antiparallel to $\vk_0(t_0)$ -- in the former case, the wave packet comes close to the Dirac cone and therefore spreads {\it after} the pulse, in the latter we have $k_0(t=0)\simeq \Delta k$ and the spreading starts in the beginning but stops after some time.

If $\Ecal\gtrsim 2 \Ecal_\tref$ the echo signal should however increase again: In this regime all initial $\vk$-modes 
are shifted to the opposite side of the Dirac cone in a time too short for the wave packet to move considerably  
-- and thus change its initial shape -- in real space.
This non-monotonic behavior is confirmed by the simulation for a $\vk_0$-parallel-to-$\Ecalb$ scenario, 
black dots in Fig.~\ref{fig:E-field-data}.  The correlator $\mathcal{C}$ is practically independent of $\Ecal$
up to $\Ecal\approx\Ecal_\tref$, where a drop is seen, followed by a recovery for $\Ecal\gtrsim 2\Ecal_\tref$.

As expected on general grounds, due to its pseudospin-independence the electric field-induced momentum shift cannot be effectively 
time-inverted by the Dirac QTM.  Nevertheless, strong echoes in real space should be possible for certain setups,
notably for narrow wave packets in $\vk$-space with $\vk_0\parallel \Ecalb$.

Here, we only investigated the effect of an electric field for linear bands, but nonlinear bands can be treated analogously. There, we expect a similar behaviour. However, the additional change of the speed $|\mathbf{v}|$ and the according change of spreading strength even for $\vk_0\parallel \Ecalb$ will most likely diminish the echo strength measured by the correlation $\mathcal{C}$ as compared to the linear case.

%%%%%%%%%%%%%%%% Magnetic field %%%%%%%%%%%%%%%%%%%%%%%%%%%%%%%%%%%%%%%%%%%%%%%%%%%%%%%%%%%%%%%%%%%%%%%%%%%%

\subsection{Homogeneous magnetic field}
\label{subsec:magnetic}
%
%%%%%%%%%%%%%%%%%%%%%%%%% FIGURE %%%%%%%%%%%%%%%%%%%%%%%%%%%%%%%%%%%%%%%%%%%%%%%%%%%%%%%%%%%%%%%%%%5
\begin{figure*}
 \centering 
     \includegraphics[width=0.7\textwidth]{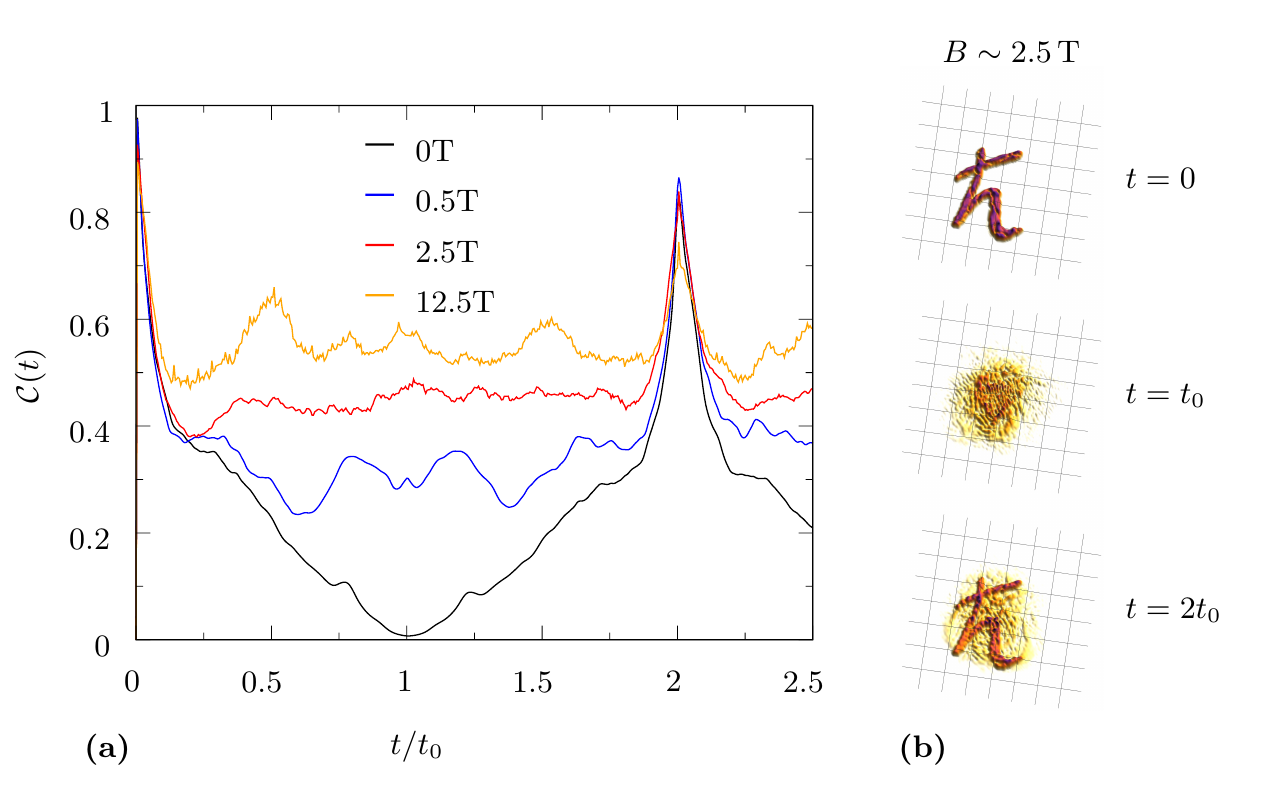}
 \centering 
\caption{
Effect of a magnetic field on the QTM of a $\hbar$-shaped wave packet. 
(a) 
Correlation $\mathcal{C}(t)$, Eq.~\eqref{eq:spaceCorr} is shown as a function of time for different magnetic field strengths ranging from $0T$ to $12.5T$. 
Although the correlation differs significantly for times $t< t_\echo = 2t_0$, the echo at $t=t_\echo$ is largely unaffected by the magnetic field. 
(b) Snapshots of the initial state ($t=0$), at the time of the pulse ($t=t_0$) and the echo ($t=2t_0$) are shown for $B\sim2.5T$. 
Although the $\hbar$-shape is completely lost at $t=t_0$, it is recovered at the echo time with a high resolution and fidelity, despite the ``noisy'' background coming from the non-reflected parts. 
}  \label{fig:Bfield}
\end{figure*}
%%%%%%%%%%%%%%%%%%%%%%%%% FIGURE %%%%%%%%%%%%%%%%%%%%%%%%%%%%%%%%%%%%%%%%%%%%%%%%%%%%%%%%%%%%%%%%%%
%
%
Does the QTM survive time reversal symmetry breaking?
Being an effective time-inversion protocol, one could expect its failure unless such a symmetry $\mathcal{T}$ 
is exactly preserved.  On the other hand the pulse Hamiltonian is itself time-dependent, $H_\pulse=H_\pulse(t)$, 
so that in general the full Hamiltonian $\mathcal{T}H\neq H\mathcal{T}$, with $\mathcal{T}$ the anti-unitary
time reversal operator.
Furthermore, the discussion from Sec.~\ref{sec:genmod} suggests that even partial time reversal symmetry, \ie limited to $H_0$,
is actually not needed: None of the QTM requirements appeal to it.
In order to explicitly check the QTM resilience to time inversion symmetry breaking, let us consider
the reference two-dimensional Dirac system of Eq.~\eqref{eq:Dirac} in the presence of an out-of-plane magnetic field:
\begin{equation}
 H = \hbar v_F \left(\vk+\frac{e}{\hbar}\mathbf{A}(\br) \right)\cdot\vsigma + f(t) M\sigma_z.
\label{eq:H_magnetic}
\end{equation}
In the symmetric gauge $\mathbf{A}(\br) = B/2 (-y,x,0)$ and following the notation of Sec.~\ref{subsec:pert:gen}, 
we have $\mathcal{V}^B(\br) =  e v_F {\bf A}\cdot\vsigma$.
Since condition \eqref{eq:Req-pert} is fulfilled with
\begin{equation}
 \sigma_z \mathcal{V}^B(\br) \sigma_z = -\mathcal{V}^B(\br),
\end{equation}
we are in case \casepertC~of the summary in Subsec.~\ref{subsec:pert:gen}.
Therefore, the magnetic field does {\it not} modify the QTM effectiveness, no matter its strength.  
This is verified numerically in Fig.~\ref{fig:Bfield}, where,
similar to Ref.~\cite{reck2017}, an $\hbar$-shaped wave packet is time-evolved.
In panel (a) the density correlator $\mathcal{C}(t)$ is plotted for different magnetic field strengths,
while panel (b) shows snapshots of the wave packet at the initial ($t=0$), pulse ($t=t_0$) and echo ($t\simeq2t_0$) times for $B=2.5$T. 
The pulse strength and length are respectively $M= 3.33 \langle E \rangle$ and $\Delta t M/\hbar\sqrt{1+\langle E \rangle^2/M^2} = \pi/2$,
with $\langle E \rangle$ the mean energy of the $\hbar$-shaped wave packet.
The latter is taken as $\langle E \rangle \simeq 100\,$meV.
%All displayed magnetic field values in Tesla correspond to a mean energy $\langle E \rangle \simeq 100\,$meV.
Due to the magnetic field the wave packet does not spread radially but undergoes some torsion.
This explains why the spatial overlaps depend on the $B$-field value and 
differ quite a lot over most of the timeline shown in panel (a). 
However at the echo time $t\simeq 2t_0$ a distinct and $B$-independent \footnote{Note that the echo strength for varying $B$-field is not truly independent, but behaves according to Eq.~\eqref{eq:B-echostrength}, where the composition of the wave packet in Landau levels depends on the $B$-field strength. What we want to express is that independent on the strength of the $B$-field, a distinct echo is possible with the QTM mechanism.} peak can be seen.
The physical reason can be explained rather easily looking at the Dirac Landau levels of the Hamiltonian
\eqref{eq:H_magnetic}.  The levels and their corresponding energies are (see \eg\cite{goerbig2011})
\begin{align}
 |\varphi^\pm_n\rangle &= \frac{1}{\sqrt{2}} \begin{pmatrix}   |n-1\rangle \\ \pm   |n\rangle            \end{pmatrix},\\
 E^\pm_n &= \pm v_F \sqrt{2\hbar eB }\; \sqrt{n},
\end{align}
with $n\in\mathbb{N}_0$ and $|n\rangle$ the harmonic oscillator eigenstates.
The transition operator is still $\sigma_z$, mapping particle- and hole-like eigenstates to one another,
\begin{equation}
 \sigma_z|\varphi^\pm_n\rangle = |\varphi^\mp_n\rangle,
\end{equation}
without changing the Landau level index $n$.
During the pulse the Landau levels are modified to 
\begin{align}
 |\chi_n^\pm\rangle &= \frac{1}{\sqrt{E_n^2+(\epsilon^\pm_n-M)^2}} \begin{pmatrix}    E^+_n |n-1\rangle \\ ( \epsilon^\pm_n -M )    |n\rangle            \end{pmatrix},\\
\epsilon^\pm_n &= \pm \sqrt{M^2+ E_n^2}.
\end{align}
For a wave packet consisting of particle-like Landau levels
\begin{equation}
 |\psi^+\rangle = \sum\limits_n c_n |\varphi^+_n\rangle,
\end{equation}
the expectation value of the Dirac velocity operator $\hat v_{x/y} = v_F \sigma_{x/y}$ 
has the opposite sign compared to its chiral (hole-like) counterpart
\begin{equation}
 |\psi^-\rangle = \sum\limits_n c_n |\varphi^-_n\rangle,
\end{equation}
that is
\begin{equation}
\langle\psi^- \mid \hat v_{x/y} \mid  \psi^-\rangle  = -  \langle \psi^+ \mid \hat v_{x/y} \mid \psi^+ \rangle .
\end{equation}
There follows that the torsions taking place before and after the pulse exactly cancel each other,
and the initial shape of the wave packet is recovered (see also Fig.~\ref{fig:Bfield}(b)) up to background noise from the parts of the wave packet which stay in the electron-like Landau levels and therefore do not invert their motion.
Quantitatively, the transition amplitude $|\varphi^\pm_n\rangle\,\leftrightarrow\,|\varphi^\mp_n\rangle$ 
can be computed along the lines of Subsec.~\ref{subsec:gentransamp}, yielding
\begin{align}
 A_n &= \langle \varphi^{\mp}_n \mid U(t_0+\Delta t,t_0 ) \mid \varphi^{\pm}_n \rangle \nonumber \\
% &= -\iu \langle \varphi^{-s}_n \mid\frac{H }{ \epsilon_n^+ } \mid \varphi^{s}_n \rangle \sin\left(\frac{\epsilon_n^+   \Delta t}{\hbar}\right) \nonumber \\
&=  \frac{-\iu }{ \sqrt{1+ E_n^2/M^2}   } \; \sin\left(\frac{M\Delta t}{\hbar} \sqrt{1+ E_n^2/M^2}\right).
\label{eq:B-transamp}
\end{align}
This is identical to the magnetic field-free expression \eqref{eq:transampl-pristgraph},
with the Landau level energy $E_n$ substituting the $\vk$-mode energy $E_\vk$ of Eq.~\eqref{eq:E-Dirac}.
For an initial wave packet
\begin{equation}
 |\Psi_0\rangle = \sum\limits_{s=\pm}\sum\limits_n \, \alpha_n^s \;| \varphi_n^s\rangle,
\end{equation}
the density correlation at the echo time reads
\begin{equation}
 \mathcal{C}(t_\echo) = \int \,{\rm d}^2\br \, \Big|\psi(\br,0)\Big| \, \Big|\sum\limits_{s=\pm}\sum\limits_n \, \alpha_n^s A_n \langle \br\mid \varphi_n^s\rangle\Big|,
\label{eq:B-echostrength}
\end{equation}
and should be compared to magnetic field-free case from Ref.~\cite{reck2017}:
\begin{equation}
 \mathcal{C}(t_\echo) =  \int \,{\rm d}^2\br \, \Big|\psi(\br,0)\Big| \, \Big|\int \frac{{\rm d}^2\vk}{(2\pi)^2}  \, A_\vk \psi(\vk,0)e^{i\vk\cdot\br}\Big|.
\end{equation}
The only difference is the sum over the Landau levels as compared to the $\vk$-integral. 
The discrete character of the sum starts to play a role when the Landau level spacing becomes comparable to the energy width 
of the wave packet.  This can be seen in Fig.~\ref{fig:Bfield}(a), where the echo strength visibly differs only for $B=12.5$T.

%%%%%%%%%%%%%%%%%%%%%%%%%%% MANY BODY BLABLABLA %%%%%%%%%%%%%%%%%%%%%%%%%%%%%%%%%%%%%

\section{Many-body effects}
\label{sec:manybody}

Our discussion of the QTM protocol did not take into account many body physics:
We assumed having a collection of non-interacting quasiparticles of unspecified nature -- fermionic or bosonic --
propagating coherently in space.  Obviously the adequacy of these assumptions depends on the actual system
used to implement the QTM, \eg topological insulators, graphene, cold atoms in optical lattices.
Quantitative statements would necessarily refer to specific situations, 
while we wish to keep our treatment broad and general.  We therefore discuss two important features of the many body problem,
namely the role of quantum statistics and that of interaction-induced decoherence, at a qualitative level only.

The first feature turns out to be irrelevant for bosons but potentially restrictive for fermions.
This is because our QTM is based on transitions between particle- and hole-like branches of the energy spectrum, 
implicitly assuming such states to be accessible.  While this is always the case for bosons,
the Pauli principle sets clear constraints for fermionic systems.
Indeed, fermionic statistics plays a role already within a non-interacting picture. 
For a direct application of the QTM concepts discussed so far, 
such a role boils down to requiring the whole energy window of the Rabi-type particle-hole oscillations
to be away from the Fermi sea -- ``hot quasiparticles'' are needed.
\footnote{Technically, a driving- and scattering-free scenario avoids such a requirement: An initial pulse
vertically excites quasiparticles out of the Fermi sea, which can then perform Rabi-like oscillations
via their own empty states created by the initial pulse.  The absence of scattering ensures that such
states will not be filled by other quasiparticles.}

The second feature represents a fundamental limitation to the QTM protocol in its present form,
no matter the nature of the quasiparticles involved.  The role of decoherence can be 
phenomenologically understood by analogy with the disordered QTM scenario from Sec.~\ref{subsec:pert:gen},
at least as long as interaction effects can be treated perturbatively.
In Sec.~\ref{subsec:pert:gen} we saw that disorder leads to dephasing of the $\vk$-mode ensemble making up a wave packet.
In the weak-disorder (perturbative) regime this causes the decay of the echo fidelity 
on a timescale of the elastic (transport) lifetime $\tau_e$ \cite{reck2017}.  
Inelastic scattering, due to \eg electron-electron or electron-phonon interactions, 
breaks each $\vk$-mode phase coherence on a timescale $\tau_i$.
The practical consequence from the point of view of an injected wave packet is however hardly distinguishable
from the disorder-induced decay of the echo fidelity -- except that the latter would now take place on the
timescale $\tau_i$, rather than $\tau_e$.    

How severe are the above statistics and decoherence constraints? 
In artificial cold atom setups they appear actually rather mild, while in standard electronic systems
they can be serious.  Yet, we emphasize that the injection and coherent propagation
(over several microns) of hot carriers in two-dimensional electron gases was recently achieved \cite{kataoka2016}.
Furthermore, it has just been shown that high-energy carriers retain their coherence when excited far
enough above the Fermi sea \cite{reiner2017}.
Moreover $\tau_i$ is in the range of $1$-$100\,$ps in graphene \cite{eless2013} and bilayer graphene \cite{engels2014} at a temperature of around $1K$,
suggesting the implementation of an electronic QTM to be feasible.

%%%%%%%%%%%%%%%%%%%%%%%%%%%%%%%%%%%%%%%%%%%%%%%%%%%%%%%%%%%%%%%%%%%%%%%%%%%%%%%%

\section{Summary}
\label{sec_conclusions}

We generalized the Dirac QTM principles \cite{reck2017} to arbitrary two-band systems.
The basic requirements are discussed in Sec.~\ref{sec:genmod} and summarized in Tab.~\ref{table_reqs}.
The band group velocity has to meet condition \reqA~therein, 
ensuring that switching from one branch to the other reverses the velocity of the initial wave packet. 
The external pulse needs instead to be chosen such that eigenstates from one band are efficiently mapped 
onto those of the other, see requirement \reqB~in Tab.~\ref{table_reqs}.
Conditions \reqC ~and \reqD~see to it that the transition amplitudes, Eq.~\eqref{eq:gen-transamp},
for modes building up a given wave packet are maximized, so that an echo with high fidelity can be achieved.

In Sec.~\ref{sec:twoband-examples} our general statements were verified for various band types: Dirac dispersion  (in the long pulse limit),
chiral-asymmetric linear bands, hyperbolic and parabolic bands.  In the analytically accessible regimes results
agree with those of numerical simulations, and show that high fidelity echoes can be achieved in all considered systems.
Possible physical realizations of the latter are direct gap semiconductors or bilayer graphene. 
Note that although all simulations have been carried out in two dimensions, the analytic calculations are as well valid for three-dimensional systems.

In Sec.~\ref{sec:Perturbations} we investigated the influence of static but inhomogeneous potentials on
the efficiency of the QTM protocol. In most situations, such potentials are  detrimental to the QTM,
but there can be remarkable exceptions.  In particular, in graphene the QTM principle is not 
affected by the presence of a strong out-of-plane magnetic field, see Fig.~\ref{fig:Bfield},
even though the latter breaks time-reversal symmetry.  On the contrary, the dynamics due 
to an additional in-plane electric field cannot in general be time-inverted by the protocol.  Nevertheless, for linear bands
distinct echoes in positions space are achieved under certain circumstances, see Fig.~\ref{fig:Efield-wavepackets}.

Finally, in Sec.~\ref{sec:manybody} we discussed many-body effects at a qualitative level,
and argued about the feasibility of experimentally implementing a QTM in different systems.
We emphasize however that our treatment is meant to be general, not aiming at realistic estimates tied
to specific realizations of the considered Hamiltonians.

To conclude, we believe that, given the rapid experimental developments in time-dependent control, multi-band systems offer rich settings for novel echo dynamics.

%%%%%%%%%%%%%%%%%%%%%%%%%%%%%%%%%%%%%%%%%%%%%%%%%%%%%%%%%%%%%%%%%%%%%%%%%%%%%%

\begin{acknowledgments}
We acknowledge support from the Deutsche Forschungsgemeinschaft within SFB 689 (project C6), SFB 1277 (project A07) and GRK 1570.
We thank Viktor Krueckl for valuable discussions and for providing TQT.
\end{acknowledgments}

%%%%%%%%%%%%%%%%%%%%%%%%%%%%%%%%%%%%%%%%%%%%%%%%%% APPENDIX %%%%%%%%%%%%%%%%%%%%%%%%%%%%%%%%%%%%%%%%%%%%%%%%%%%%%%555
\appendix

\section{Derivation of the echo time}
\label{app:t_echo}

To calculate the echo time, we need to have a look at the full propagation. We will first focus on the phase of a single eigenstate of ${H}_0$ that is switched to the opposite band by the pulse.
Since the $\vk$-dependent part of this phase yields the translation of the given $\vk$-modes, it is supposed to vanish in the case of an echo.

The phase acquired until the pulse at $t_0$ is given by
\begin{equation}
 \me^{-\frac{\iu}{\hbar} {H}_0 t_0} \mid \varphi^s_\vk \rangle =  \me^{-\frac{\iu}{\hbar} E_s(\vk) t_0} \mid \varphi^s_\vk \rangle.
\end{equation}
After the pulse at $t=t_0+\Delta t$, the band-inverted state is given by the transition amplitude of Eq.~\eqref{eq:gen-transamp}.
\begin{align}
 &\me^{-\frac{\iu}{\hbar} E_s(\vk) t_0} \mid \varphi^s_\vk \rangle \xlongrightarrow{t_0+\Delta t} \nonumber \\
\me^{-\frac{\iu}{\hbar} E_s(\vk) t_0}
(-\iu)\frac{ h_1^\perp}{ \hbar\Omega_\vk  } &\sin(\Omega_\vk \Delta t) \me^{-\frac{\iu}{\hbar} \Delta t  (h^0_0 + h^0_1)} \mid \varphi^{-s}_\vk \rangle 
\end{align}
Note that the ongoing state, which was not band-inverted, is omitted here, since it will not contribute to the echo anyway.
Since we are interested in the $\vk$-dependent phase, we will omit the factor $(-\iu)\frac{ h_1^\perp}{ \hbar\Omega_\vk  } \sin(\Omega_\vk \Delta t)$. However, in Subsec.~\ref{subsubsec:verylongdt}, we show that the sine term leads to a splitting of the wave packet for long $\Delta t$ and therefore there will be two echoes. 
In that sense, we derive the mean time of these two splitted echoes, which is for short $\Delta t$ essentially the same.

The phase at the echo time $t_\echo = t_0+\Delta t + t_1$ is again given by
\begin{equation}
\exp\left(-\frac{\iu}{\hbar} \left(E_s(\vk) t_0 + \Delta t (h^0_0 + h^0_1) + E_{-s}(\vk) t_1\right)\right) \mid \varphi^{-s}_\vk \rangle ,
\label{eq:app-phases}
\end{equation}
where we have to identify $t_1$ where the phase is $\vk$-independent. 
This $\vk$-independence leads to the relation between the energies of Eq.~\eqref{eq:velrequirement}, which was based on physical intuition.

Moreover, we see that the $\vk$-dependent term of $(h_0^0+ h_1^1)$ has to be the same as of $E_s(\vk)$, or equivalently
\begin{equation}
\frac{\partial}{\partial k} (h_0^0+ h_1^1)  = \xi_h \frac{\partial}{\partial k} E_s(\vk), \quad \xi_h \in \mathbb{R},.
\label{eq:h_0req}
\end{equation}
Otherwise, different modes will return to the initial position at different times, and thus the echo would be washed out.
Assumption \eqref{eq:h_0req} seems hard to be fulfilled. Nevertheless for the examples shown in the main text, it works even in a linear (parabolic) band structure 
with different slopes (curvatures) of positive and negative bands, due to the monomial $k$-dependence.

If Eq.~\eqref{eq:h_0req} is fulfilled, $t_1$ yields 
\begin{equation}
 t_1 = \left(t_0 + \xi_h \Delta t\right)\xi_v
\end{equation}
and the echo time becomes
\begin{equation}
 t_\echo = \left(1 + \xi_v\right) t_0 + \left( 1+  \xi_h\xi_v \right) \Delta t.
\end{equation}

\section{Echo time - long pulse durations}
\label{app:t_echo-longDeltat}

The process of the separation of the individual sub-wave packets in the long $\Delta t$-limit can be investigated by considering the phase similar to App.~\ref{app:t_echo}. 
There, the kinetic phases which are accumulated during the propagation are carefully examined. 
Since $\vk$-dependent phases of the form $\me^{\iu \vk\cdot \br_0}$ lead to a translation in real space by  $\br_0$ due to the properties of the Fourier transformation, the echo happens when all $\vk$-dependent phases cancel. 

For convenience, we consider an initial wave packet $\psi_0(\br)$ with $\langle\hat{\br}\rangle = 0$, which is peaked in $\vk$-space around $\vk_0$ with width $\Delta k$, \eg a Gaussian wave packet.
At some time $t^\prime = t_0+\Delta t + t_1^\prime$ after the pulse, the amplitude of each initial mode $| \varphi^{s}_\vk \rangle$ which has undergone a band transition to $| \varphi^{-s}_\vk \rangle$ reads
\begin{align}
\langle \varphi^{-s}_\vk \mid U(t^\prime, &0 )\mid  \varphi^{s}_\vk \rangle \nonumber \\
&= A_\vk \exp\left(-\frac{\iu}{\hbar} \left(E_s(\vk) t_0 + E_{-s}(\vk) t_1^\prime\right)\right) .
\end{align}
We consider here only the sine of the transition amplitude of Eq.~\eqref{eq:transampl-pristgraph}, because it is the only term that adds to the $\vk$-dependent phase, and expand it in terms of exponentials:
\begin{equation}
  \sum\limits_{l=\pm1} \frac{l}{2}\me^{\iu l \Omega_\vk\Delta t}  \exp\left(-\iu v_F k  \left(t_0 -t_1^\prime\right)\right)
\end{equation}
From now on, we assume that $\Omega_\vk$ = $\Omega_k$, \ie independent of the direction of $\vk$ which is true in the case of graphene with a pulse that opens a mass gap, for instance. 

Since the $k$-linear term is responsible for a translation, we expand $\Omega_k$ in $k$ around the value $|\vk_0|=k_0$ where the wave packet is peaked, 
{\small 
\begin{equation}
\sum\limits_{l=\pm1}  \frac{l}{2}\, \exp\left(-\iu v_F k  \left(t_0 -t_1^\prime - l \Delta t \left.\frac{\partial \Omega_k}{\partial k} \right|_{k_0} \right)\right) + \mathcal{O}(\Delta k)^2,
\label{eq:phase-linear-pristgraph}
\end{equation}
}
omitting again all constant phase terms. 
The translation is zero (echo), when
\begin{equation}
 t_1^\prime = t_0 \pm \Delta t  \left.\frac{\partial \Omega_k}{\partial k} \right|_{k_0}.
\end{equation}
For graphene with a mass pulse as in Subsec.~\ref{subsubsec:verylongdt}, we have 
$\Omega_{k_0}$\;=\;$M/\hbar \sqrt{1+\kappa_0^2}$ with $\kappa_0 $\;=\;$ \hbar v_F k_0/M$, and the echo time simplifies to
\begin{equation}
 t_1^\prime = t_0 \pm \Delta t  \frac{\kappa_0}{\sqrt{1+\kappa_0^2}} = t_0\pm t_2.
\end{equation}
Thus, the peaks of the wave packets return at distinct times  
\begin{equation}
 t^\pm_\peak = 2t_0 +\Delta t \left(1\pm   \frac{\kappa_0}{\sqrt{1+\kappa_0^2}}\right). 
\end{equation}
Note that in Eq.~\eqref{eq:phase-linear-pristgraph}, the quadratic (and possibly higher-order) terms in $\Delta k$ lead in principle to a spreading as in the free Schr\"odinger equation, which will in general diminish the echo - the longer $\Delta t$ the weaker the echo. 
However, we see in the simulations of Subsec.~\ref{subsubsec:verylongdt} that the spreading does not play a role for the used $\vk$-width.

\section{Expansion of the time evolution operator}
\label{app:expansion-time-evo-pulse}
We consider a system as defined in Subsec.~\ref{subsec:pert:gen} with a static Hamiltonian
\begin{equation}
 H_0+\mathcal{V} = h_0(\vk) \hbn_\vk\cdot \vsigma +  V_l(\vq)\hbl_\vq\cdot \vsigma
\end{equation}
that has eigenenergies $\epsilon^\pm_n$ and eigenstates $|\phi_n^\pm\rangle$, and a pulse Hamiltonian
\begin{equation}
 H_\pulse = M \hbm\cdot\vsigma,
\end{equation}
with $\hbn_\vk, \hbl_\vq \perp \hbm, \forall \vk,\vq$.
The expansion of the time-evolution operator acting during the pulse reads
\begin{align}
 \exp&\left(-\frac{\iu}{\hbar} (H_0+\mathcal{V} + M \hbm\cdot\vsigma)\Delta t \right) \nn \\
&= \sum\limits_{j=0}^\infty \frac{1}{j!} \left(-\frac{\iu\Delta t }{\hbar}\right)^j  (H_0+\mathcal{V} + M \hbm\cdot\vsigma)^j.
\label{eq:timeEvopulse-expansion0}
\end{align}
To simplify the time-evolution operator, we make use of the anticommutation relations of Pauli matrices in orthogonal directions,
\begin{equation}
 \lbrace \hbn_\vk\cdot\vsigma, \hbm\cdot\vsigma \rbrace = 0 =  \lbrace \hbl_\vq\cdot\vsigma, \hbm\cdot\vsigma \rbrace
\end{equation}
and use the fact that $M$ is a real number that commutes with $H_0(\vk)$ and $\mathcal{V}(\vq)$.
With that, we see that even and odd powers of the Hamiltonians in Eq.~\eqref{eq:timeEvopulse-expansion0} become
\begin{align}
 (H_0+ &\mathcal{V}+ M \hbm\cdot\vsigma)^j  \nn \\
&= \left\lbrace\begin{array}{ll} \sqrt{(H_0+\mathcal{V})^2+M^2}^j, & \text{j even,} \\ \sqrt{(H_0+\mathcal{V})^2+M^2}^j \frac{(H_0+\mathcal{V}) + H_\pulse}{\sqrt{(H_0+\mathcal{V})^2+M^2}}, & \text{j odd,}\end{array}\right.
\end{align}
which is why we can expand Eq.~\eqref{eq:timeEvopulse-expansion0} into sine and cosine terms:
{\small 
\begin{widetext}
\begin{align}
\exp\left\lbrace -\frac{\iu}{\hbar} (H_0+\mathcal{V} + M \hbm\cdot\vsigma)\Delta t \right\rbrace
= \mathbbm{1}\cos\left(\frac{\Delta t}{\hbar}\sqrt{(H_0+\mathcal{V})^2 + M^2 }\right) 
-\iu \frac{(H_0+\mathcal{V}) + H_\pulse}{\sqrt{(H_0+\mathcal{V})^2+M^2}}\sin\left(\frac{\Delta t}{\hbar}\sqrt{(H_0+\mathcal{V})^2 + M^2 }\right).
\label{eq:timeEvopulse-expansion1}
\end{align}
\end{widetext}
}
When acting on an eigenstate $|\phi_n^\pm\rangle$, we therefore get
\begin{align}
 \me^{-\frac{\iu}{\hbar} (H_0+\mathcal{V} + M \hbm\cdot\vsigma)\Delta t} \;|\phi_n^\pm\rangle
= B_n^\pm  \; |\phi_n^\pm\rangle + A_n  \;|\phi_n^\mp\rangle
\label{eq:timeEvopulse-expansion2}
\end{align}
with the amplitudes
\begin{align}
 B_n^\pm =& \cos\left(\frac{\Delta t}{\hbar}\sqrt{\epsilon_n^2 + M^2 }\right) \nn\\
 &-\iu \frac{\epsilon_n^\pm}{\sqrt{\epsilon_n^2+M^2}}\sin\left(\frac{\Delta t}{\hbar}\sqrt{\epsilon_n^2 + M^2 }\right)
\end{align}
and 
\begin{align}
 A_n =  \frac{-\iu}{\sqrt{1+\epsilon_n^2/M^2}}&\sin\left(\frac{\Delta t M }{\hbar}\sqrt{1+\epsilon_n^2/M^2 }\right).
\end{align}

% 
% 
% \bibliography{biblio_QTM_bands}

%merlin.mbs apsrev4-1.bst 2010-07-25 4.21a (PWD, AO, DPC) hacked
%Control: key (0)
%Control: author (0) dotless jnrlst
%Control: editor formatted (1) identically to author
%Control: production of article title (0) allowed
%Control: page (1) range
%Control: year (0) verbatim
%Control: production of eprint (0) enabled
%

\end{document}